\documentclass{aa}  

\usepackage{graphicx}
\usepackage{txfonts}
\usepackage{amsmath}    
\usepackage{amssymb}    
\usepackage{multirow,makecell}
\usepackage{tabularx}
\usepackage[style=base]{caption}
\usepackage{booktabs}
\usepackage{threeparttable}
\usepackage{enumitem}
\usepackage{float}
\usepackage{placeins}
\newcommand{\pcm}{\,cm$^{-2}$}  
\newcommand{\pcmc}{\,cm$^{-3}$} 
\newcommand{\msol}{\,M$_\odot$} 
\usepackage[]{hyperref}
\begin{document}

   \title{Hunting for hot corinos and WCCC sources in the OMC-2/3 filament}

   \subtitle{}

   \author{M. Bouvier
           \inst{1}\fnmsep
          \and
          A. L\'opez-Sepulcre\inst{1,2}
          \and
          C. Ceccarelli\inst{1,3}
          \and
          C. Kahane\inst{1}
          \and
          M. Imai\inst{4}
          \and
          N. Sakai\inst{5}
          \and
          S. Yamamoto\inst{4,6}
          \and
          P. J. Dagdigian\inst{7}
          }

   \institute{Univ. Grenoble Alpes, CNRS, Institut de Planétologie et d'Astrophysique de Grenoble (IPAG), 38000 Grenoble, France\\
              \email{mathilde.bouvier@univ-grenoble-alpes.fr}
         \and
             Institut de Radioastronomie Millim\'etrique (IRAM), 300 rue de la Piscine, 38406 Saint-Martin-D'H\`eres, France
         \and 
            CNRS, IPAG, F-38000 Grenoble, France
        \and
            Department of Physics, The University of Tokyo, 7-3-1, Hongo, Bunkyo-ku, Tokyo 113-0033, Japan
        \and 
            RIKEN, Cluster for Pioneering Research, 2-1, Hirosawa, Wako-shi, Saitama 351-0198, Japan
        \and    
            Research Center for the Early Unvierse, The University of Tokyo, 7-3-1, Hongo, Bunkyo-ku, Tokyo 113-0033, Japan
        \and
            Department of Chemistry, The Johns Hopkins University, Baltimore, MD 21218-2685, USA
             }

   \date{Received XXX; accepted YYY}

 
  \abstract
   {Solar-like protostars are known to be chemically rich, but it is not yet clear how much their chemical composition can vary and why. So far, two chemically distinct types of Solar-like protostars have been identified: hot corinos, which are enriched in interstellar Complex Organic Molecules (iCOMs), such as methanol (CH$_3$OH) or dimethyl ether (CH$_3$OCH$_3$), and Warm Carbon Chain Chemistry (WCCC) objects, which are enriched in carbon chain molecules, such as butadiynyl (C$_4$H) or ethynyl radical (CCH). However, none of these have been studied so far in environments similar to that in which our Sun was born, that is, one that is close to massive stars.  }
   {In this work, we search for hot corinos and WCCC objects in the closest analogue to the Sun's birth environment, the Orion Molecular Cloud 2/3 (OMC-2/3) filament located in the Orion A molecular cloud.}
   {We obtained single-dish observations of CCH and  CH$_3$OH line emission towards nine Solar-like protostars in this region. As in other, similar studies of late, we used the [CCH]/[CH$_3$OH] abundance ratio in order to determine the chemical nature of our protostar sample. }
   {Unexpectedly, we found that the observed methanol and ethynyl radical emission (over a few thousands au scale) does not seem to originate from the protostars but rather from the parental cloud and its photo-dissociation region, illuminated by the OB stars of the region. }
   {Our results strongly suggest that  caution should be taken before using [CCH]/[CH$_3$OH] from single-dish observations as an indicator of the protostellar chemical nature and that there is a need for other tracers or high angular resolution observations for probing the inner protostellar layers.}


   \maketitle
%

\section{Introduction}\label{sec:intro}

A key aspect of the chemical richness of the protostellar stage is the diversity found among Solar-like protostars. Indeed, two chemically distinct types of Solar-like protostars have been identified. On the one hand, hot corinos \citep{ceccarelli2000,ceccarelli2007} are compact ($<$ 100 au), dense ($>$ 10$^7$\ \pcmc), and hot ($>$ 100 K) regions, enriched in interstellar Complex Organic Molecules (hereafter iCOMs; for example, CH$_3$OH, CH$_3$CHO, HCOOCH$_3$; \citealt{herbst2009}; \citealt{ceccarelli2017}). On the other hand, Warm Carbon Chain Chemistry (hereafter WCCC; \citealt{sakai2008b}, \citealt{sakai2013}) objects have an inner region deficient in iCOMs but a large ($\approx $2000 au) zone enriched in carbon chain molecules (e.g. CCH, c-C$_3$H$_2$, C$_4$H). This dichotomy does not seem to be absolute as at least one source, the protostar L483, presents both hot corino and WCCC characteristics \citep{oya2017}. Understanding what causes this chemical diversity is a fundamental step in understanding the formation and the evolution of a planetary system like our own and, perhaps, to understand the appearance of life on Earth. Furthermore, given our Sun was formerly a protostar, it is natural to consider whether it may have experienced a hot corino phase, a WCCC phase, or neither of the two in its youth.  In this respect, we recall that the Solar System formed in a large stellar cluster in proximity to high-mass stars (M$_{*}$ $\geq$ 8\ \msol; \citealt{adams2010}, \citealt{pfalzner2015}).

\begin{table*}
\makegapedcells
\setcellgapes{6pt}
\caption{Summary of the excitation temperatures (T$_{\text{ex}}$) and column densities (N$_{\text{tot}}$) of the different tracers of hot corinos (CH$_3$OH) and WCCC objects (CCH and C$_4$H) derived in previous surveys obtained with single-dish telescope observations.}
\label{tab:surveysold}
\resizebox{\textwidth}{!}{
\begin{threeparttable}
\begin{tabular}{|c|c|c|c|c|c|c|c|c|c|c|c|}
\hline
\multirow{4}{*}{Object type}& \multirow{4}{*}{Molecule}& \multicolumn{2}{c|}{Graninger et al. 2016 }& \multicolumn{2}{c|}{Lindberg et al. 2016} & \multicolumn{2}{c|}{Higuchi et al. 2018 }&\multicolumn{2}{c|}{L1527 \tnote{b}} &\multicolumn{2}{c|}{ IRAS 16293-2422 \tnote{c}}\\
\cline{3-12}
&& T$_{\text{ex}}$ & N$_{\text{tot}}$& T$_{\text{ex}}$ & N$_{\text{tot}}$& T$_{\text{ex}}$ & N$_{\text{tot}}$&T$_{\text{ex}}$ & N$_{\text{tot}}$&T$_{\text{ex}}$ & N$_{\text{tot}}$\\
&&[K] & [$\times 10^{13}$cm$^{-2}$]& [K] &[$\times 10^{13}$cm$^{-2}$] &[K] &[$\times 10^{13}$cm$^{-2}$] &[K] &[$\times 10^{13}$cm$^{-2}$]&[K] &[$\times 10^{13}$cm$^{-2}$]\\
\hline
 \multirow{2}{*}{WCCC tracers}&CCH&...&...&...&...&8 $-$ 21&$\leq 52$&8 $\pm$ 1&33 $\pm$ 3&18 $\pm$ 6&9 $\pm $ 6\\
 \cline{2-12}
 &C$_4$H&7 $-$ 15&$<4$&7 $-$ 16.5\tnote{a} &$<1$&...&...&12.3&20 $\pm$ 4&12.3&1.2 $\pm$ 0.3\\
 \hline
Hot corino tracer&CH$_3$OH&4 $-$ 7&0.2 $-$ 11&$<123$\tnote{a}&0.2 $-$ 95&8 $-$ 21&0.5 $-$ 16&8 $\pm$ 1&8 $\pm$ 1&84 $\pm$ 6&90 $\pm $ 10\\
\hline
\multicolumn{2}{|c|}{Resolution (au)} & \multicolumn{2}{c|}{4100 - 13000} & \multicolumn{2}{c|}{7800 - 8200}  & \multicolumn{2}{c|}{2400 - 4900}  & \multicolumn{2}{c|}{2400 - 4900}  & \multicolumn{2}{c|}{500 - 2000}  \\
\hline
\end{tabular}
\tablefoot{
\tablefoottext{a} {Excitation temperatures taken from c-C$_3$H$_2$ and H$_2$CO APEX observations for C$_4$H and CH$_3$OH respectively \citep{lindberg2016} }
\tablefoottext{b} { \citet{higuchi2018}, \citet{sakai2008b}}
\tablefoottext{c}{\citet{vandishoeck1995}, \citet{sakai2009a}}
}
\end{threeparttable}
}
\end{table*}
So far, there have been only a few hot corinos and WCCC objects identified and almost all of them are located in low-mass star forming regions (e.g. \citealt{cazaux2003,sakai2008b,taquet2015}). Besides the relatively small number of WCCC objects in low-mass star-forming regions, WCCC characteristics have also been seen in other environments, such as the starless core L1489 \citep{Wu2019} and in the giant HII region NGC 3576 \citep{Saul2015}. A hot corino was also found in the high-mass star-forming region of Orion, HH212-MM1 \citep{codella2016}, but as with the other hot corinos, it is located far from massive stars. Previous observational studies have been carried out towards hot corinos and WCCC objects in an effort to better understand them. Some studies have been targeted on just one type of object (either hot corino or WCCC object; e.g. ~\citealt{Caux2011, jorgensen2016, lopez2017, oya2017, ospina2018, agundez2019, bianchi2019, yoshida2019}), or one type of molecular tracer (Carbon-chain or iCOMs; e.g. \citealt{law2018, Wu2019} ) whereas other studies targeted all kinds of protostars and selected tracers to evaluate their chemical nature (i.e. comparative statistical studies of chemical diversity; e.g. \citealt{graninger2016}, \citealt{lindberg2016}, \citealt{higuchi2018}). For the latter, the  method employed is based on observations with single-dish telescopes (scale of $\approx 10,000$ au) of small carbon chains (e.g. CCH, C$_4$H) and methanol (CH$_3$OH) as tracers of WCCC objects and hot corinos, respectively. 
However, this method presents some caveats. Indeed, emission of small carbon chains are usually extended and present in the Photo-Dissociation Regions (hereafter PDRs) surrounding the molecular clouds (e.g. \citealt{pety2005,cuadrado2015}). As for CH$_3$OH, it is also a species that has been found in PDRs as well as in molecular clouds (e.g. \citealt{leurini2010,guzman2013,cuadrado2017, punanova2018}). Thus, an important contribution from the parental molecular cloud  or from the surrounding PDR may occur when observing those molecules with single-dish telescopes. \\
\indent In this context, the goal of the present work is twofold: (1) to identify the nature of several protostars, hot corinos, or WCCC objects in a region containing high-mass stars and whether it depends on the object position in the cloud; this will help us to understand whether the Sun passed through a hot corino or a WCCC object phase. Indeed, if only hot corinos are found in Orion Molecular Cloud 2/3 (OMC-2/3), this would strongly suggest that our Sun also underwent a hot corino phase during its youth and vice versa; (2) to verify the reliability of using single-dish observations of small hydrocarbons and methanol to classify the chemical nature of the protostars.\\
\indent To reach these two goals, we obtained new IRAM-30m and Nobeyama-45m observations of ethynyl radical (CCH) and methanol towards a sample of nine known protostars in the closest high- and low- mass star forming region, OMC-2/3, and a map of a portion of it. Following \citet{higuchi2018}, we used the [CCH]/[CH$_3$OH] abundance ratio to make a first assessment of the chemical nature of the targeted sources: a small ($\leq$ 0.5) ratio would be suggestive of a hot corino candidate whereas a large ($\geq$ 2) ratio would rather be suggestive of a WCCC candidate, as we will discuss in detail in Section \ref{sec:targets}. Moreover, this abundance ratio is about one order of magnitude different when comparing the hot corino and WCCC templates sources (see Tab.\ref{tab:surveysold}), which would a priori justify the use of this initial criterion. \\
\indent This paper is structured as follows. In Section \ref{sec:previous-survesy}, we briefly review the previous surveys with single-dish observations aimed to study the chemical nature of low-mass protostars. In Section \ref{sec:targets}, we briefly describe the OMC-2/3 region, the selected source sample and explain the choice of the molecular species that we targeted. A description of the observations are presented in Section \ref{sec:observations}. In Section \ref{sec:results}, we show the results of the analysis of the observed lines. In Section \ref{sec:phys-parameters}, we present the derived physical conditions (temperature and density) of the gas which emits the detected CCH and CH$_3$OH lines and their column densities. In Section \ref{sec:discussio}, we discuss our findings. We finally end with some concluding remarks in Section \ref{sec:conclusion}.

\section{Prior surveys of protostellar chemical diversity with carbon chain-iCOM ratios}\label{sec:previous-survesy}
The first efforts aimed at improving statistics on the chemical nature and diversity of low-mass protostars were only recently begun. The chemical tool used is the abundance ratio between carbon chains (e.g. CCH, C$_4$H) and iCOMs (e.g. CH$_3$OH). Those families of molecules are abundant in WCCC objects and in hot corinos, respectively. We present in this section only surveys that used the chemical tool cited above to identify hot corinos and WCCC sources. 

\citet{graninger2016} investigated the relationship between C$_4$H and CH$_3$OH in 16 embedded protostars located in the northern hemisphere. The abundance ratio [C$_4$H]/[CH$_3$OH] for a typical WCCC is 2.5 and for typical hot corinos (IRAS16293-2422, IRAS 4A, IRAS 4B, Serpens MMS4), $<0.15$. With this definition, among their 16 sources, there would be five WCCC sources and at most two hot corinos. However, the temperatures derived are likely too low ($\leq$ 15 K) to correspond to the lukewarm envelope of the protostars. Their main results are: 1) there is a positive correlation between the column densities of C$_4$H and CH$_3$OH, indicating that the two species are present in a lukewarm environment in the protostellar envelopes; and 2) they found a lower amount of CH$_3$OH than in hot corinos and a lower amount C$_4$H than in WCCC sources. The under-abundance of carbon chains in this source sample has been confirmed by \citet{law2018}.

\indent Similarly, \citet{lindberg2016}, observed 16 low-mass protostars in the southern hemisphere and, including the results from \citet{graninger2016}, they investigated the origin of C$_4$H and CH$_3$OH in the protostellar envelopes. Using the same abundance ratio, two sources would be labelled as WCCC protostars and about six as hot corinos. Contrary to  \citet{graninger2016}, they did not observe evidence of a correlation between the column densities of the two species and they concluded that CH$_3$OH would reside in the warmer inner regions of the protostellar environment, whereas C$_4$H would reside rather in the cooler outer regions of the protostellar environment. However, the derived excitation temperature for  CH$_3$OH never exceeds 36 K, except in the well-known hot corino IRAS 16293-2422. Those temperatures seem too low for CH$_3$OH to originate in the hot corinos. Indeed, for hot corinos (T$\geq 100$ K; \citealt{ceccarelli2007}), we expect excitation temperatures of at least 50 K to support an origin of emission dominated by the hot corino. \\
\indent Finally,  \citet{higuchi2018}, performed a survey of 36 low-mass  protostars in the Perseus region. They compared the abundance ratio $\left[\text{CCH}\right]/ \left[\text{CH$_3$OH}\right]$ of each source to that of L1527, the prototypical WCCC source, to characterise their chemical nature. From this criterion, at most four sources can be WCCC protostars and 14 sources seem to be hot corinos. Two main results of their work are: 1) the majority of the sources have intermediate chemical composition between hot corinos and WCCC types; and 2) WCCC objects tend to be found at the edge of molecular clouds or in relative isolation, whereas the hot corinos tend to be located in the centre of molecular clouds.  Also in this case, the derived excitation temperatures are low (in the range of 8 - 21 K), so it is not entirely clear whether the observed molecular emission is contaminated by the molecular cloud.\\
\indent Table.~\ref{tab:surveysold} summarises the derived excitation temperatures and column density of CCH, C$_4$H and CH$_3$OH in the three surveys, as well as those derived for L1527 (the WCCC prototype) and IRAS 16293-2422 (the hot corino prototype). Based on these surveys and their definition of hot corinos and WCCC objects, about 33\% of the low-mass protostars are hot corinos, 17\% can be classified as WCCC sources and the rest (50\%) do not seem clearly to belong to any of the two categories. We note that the definition of hot corino and WCCC object varies from one survey to another.  Overall, 66 low-mass protostars have been surveyed upon adding up all of these studies, but all of them concern regions with no massive stars nearby, which is basis of the next step and overall goal of our study.

\begin{figure}[H]
        \resizebox{\hsize}{!}{\includegraphics{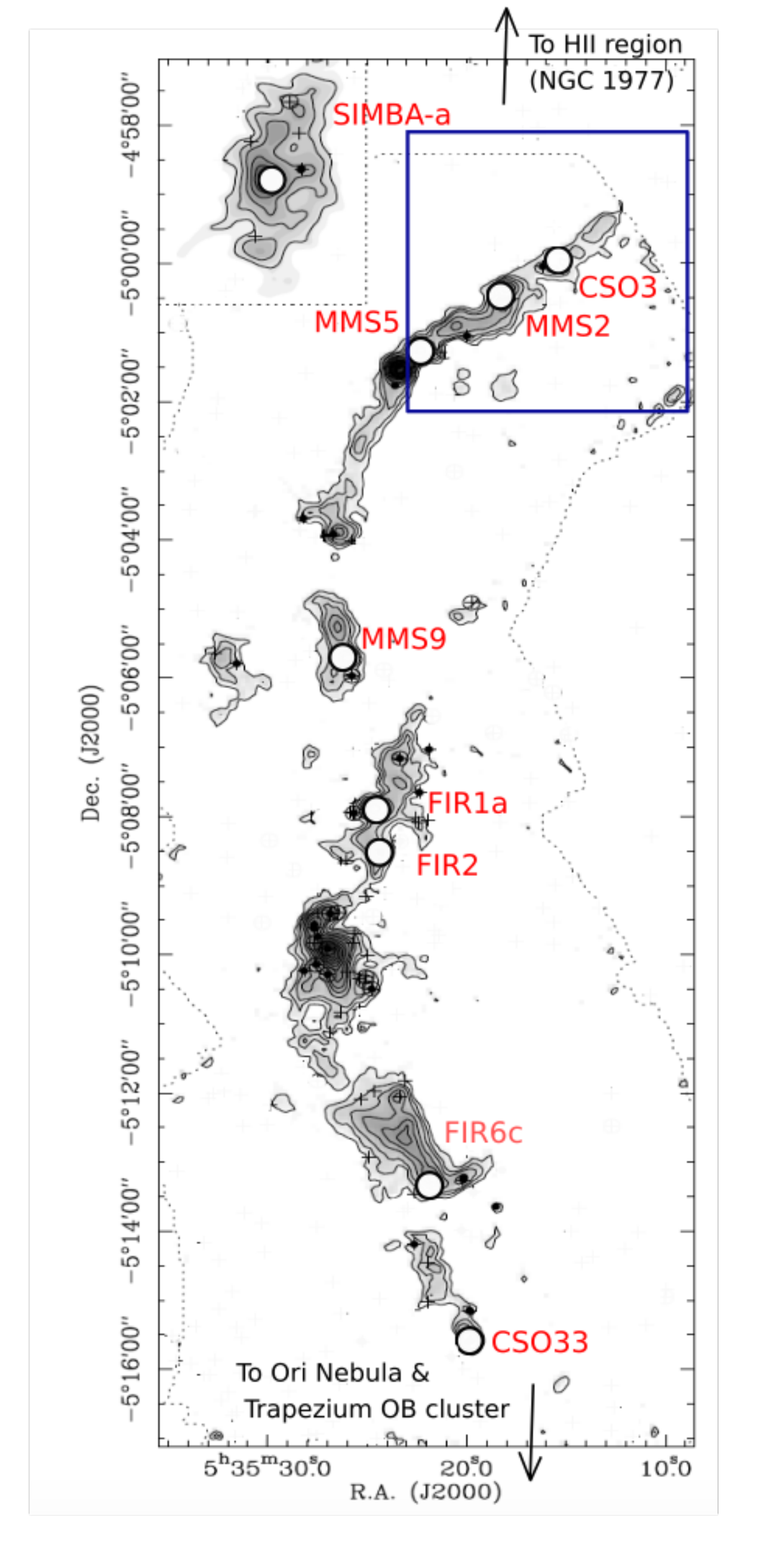}}
   \caption{Map of the continuum at 1.3mm towards the OMC-2/3 filament, adapted from \citealt{chini1997} and \citealt{nielbock2003}:The upper left insert (square grid) is taken from the SIMBA map of \citet{nielbock2003} and added on the 1.3 mm map of \citet{chini1997} whose limits are represented by the dotted lines. Contours rise linearly from a $3\sigma$ rms noise level \citep{nielbock2003}, the rms noise of the map being 25 mJy \citep{chini1997}. The selected targets of the present study are marked with white circles, which have the size of the IRAM-30m beam at 1.3mm. The names of the sources are reported in red (see Tab. \ref{tab:sources}). Please note that some black crosses and filled dots are present throughout the figure. They represent some of the 2MASS sources and MIR sources from TIMMI 2 observed by \citet{nielbock2003}.  The blue square indicates the region mapped with IRAM-30m at 3mm. The two arrows show the direction of the HII region NGC 1977 (top) and the Ori Nebula and Trapezium OB cluster (bottom).}
\label{fig:omc23}
\end{figure}

\section{Targeted sources and molecular tracers}\label{sec:targets}
\begin{table*}
\centering
\resizebox{\textwidth}{!}{
\begin{threeparttable}
\caption{List of the selected sources and their properties. The envelope mass of the protostars are taken from \citet{li2013} and \citet{takahashi2008}. The data for the radio component are taken from VLA cm-radio emission \citep{reipurth1999}. The data on the outflows are from \citet{takahashi2008}.}
\label{tab:sources}
\begin{tabular}{|p{1.5cm}|cccccl|c|}
\hline
\multirowthead{2}{Source}  & R.A. & Dec. & Mass  & Radio    & Outflow & Notes& \multirowthead{2}{References}\\
& J2000& J2000 &  \ \msol &(Y/N)&(Y/N)&& \\ 
\hline
CSO33 & 05:35:19.50& $-$05:15:35.0&  6 & N & Y? &  near HII region (M42)&1\\
FIR6c \tnote{a} & 05:35:21.60& $-$05:13:14.0&  5 - 9& N& Y &  Class 0 &2\\
FIR2  \tnote{a}& 05:35:24.40& $-$05:08:34.0&  4 - 10& N & Y & Class I &2\\
FIR1a & 05:35:24.40& $-$05:07:53.0&  12& Y & N & Class 0?&2 \\
MMS9 & 05:35:26.20& $-$05:05:44.0&  4.7 &Y & Y &  Class 0&2 \\
MMS5 & 05:35:22.50& $-$05:01:15.0&  4.5& N & Y &  Class 0, part of the OTF map&2 \\
MMS2 & 05:35:18.50& $-$05:00:30.0&  2.8& Y& Y &   Class I, binary , part of the OTF map&2 \\
CSO3 & 05:35:15.80& -04:59:59.0&  12 & N& N & Binary, part of the OTF map&1\\
SIMBA-a \tnote{b}& 05:35:29.80& -04:58:47.0&  6.5& ...  & Y &  Class 0, isolated, near HII region (NGC 1977) &3\\
\hline
\end{tabular}
\tablebib{
(1)~\citet{lis1998}; (2) \citet{chini1997}; (3) \citet{nielbock2003}}
\tablefoot{
 \tablefoottext{a}{Left out of the analysis due to prominent non-Gaussian line wings}
 \tablefoottext{b}{Data from the 3mm only, as there was emission contamination from the wobbler off position for the 1mm data set.}
 }
 \end{threeparttable}
 }
\end{table*}
%
\subsection{Target selection}

OMC-2/3 is part of the Orion A molecular complex and is composed of two major molecular clouds, OMC-2 and OMC-3. Those two clouds form a filament-shaped region located between the Trapezium OB stellar cluster and the NGC1977 HII region, at a distance of (393$\pm$25) pc from the Sun \citep{grosschedl2018}. 
OMC-2 lies to the north of OMC-1, which is the molecular cloud associated with the Orion Nebula, and was identified by \citet{gatley1974}. The cloud OMC-3, located further north, was labelled later by \citet{chini1997}. 
OMC-2/3 has been extensively studied since its discovery and multiple protostars, in particular Class 0 protostars, have been identified (e.g \citealt{chini1997}), as well as multiple molecular outflows (e.g \citealt{reipurth1999, williams2000, takahashi2008}), and H$_{2}$ shocks and jets \citep{yu1997}. 

What makes OMC-2/3 particularly interesting is that it is the nearest region containing both low- and high-mass forming stars. In addition, it is surrounded by bright OB stars which create HII regions and PDR. Thus, OMC-2/3 is unique target for the study of a region similar to the one in which the Solar System was born, as well as to probe the effect of different environments. For this reason we carried out a systematic study of several sources with the goal of characterising their chemical nature, that is, hot corino versus WCCC objects, and how this depends on the location of the source within OMC-2/3 and its immediate surroundings.

With this in mind,  we selected a sample of low-mass protostellar sources which satisfy the following three criteria: (1) detection in the (sub-)mm continuum emission; (2) estimation (based on the continuum) masses $\leq $12\ \msol; (3) bona fide Class 0 and I protostars (which excludes pre-stellar cores that may also be present in this region).
The resulting sample consists of nine sources, whose coordinates as well as some of their known properties, are listed in Table~\ref{tab:sources}. 
Their masses (dust+gas) cover from 2.8 to 12 \msol \citep{takahashi2008, li2013}, constituting, therefore, a rather homogeneous sample in this respect. Most of the sources are known to possess outflows, which is in line with their protostellar nature, and one third of them have detected cm-radio emission \citep{reipurth1999}, which is again a sign of a protostellar nature (e.g. \citealt{angl1995, anglada1996}). Finally, Fig. \ref{fig:omc23} shows the distribution along the OMC-2/3 filament of the selected targets: they cover the whole filament from the north end (CSO3 source) to the south one (CSO33 source). Some of them are close to the border of the molecular cloud whereas others lie well inside, where the UV radiation field from the nearby OB stars is expected to be as much as 3 orders of magnitude lower than at the edge of the filament (see \citealt{lopez2013}). This will thus allow us, as we aim, to probe different environments across the filament.
We note that we did not add to the study the famous FIR4, the brightest sub-mm source in the OMC-2 cloud, because it is already the focus of several past and future studies (e.g. \citealt{lopez2013}; \citealt{Kama2013}; \citealt{ceccarelli2014}; \citealt{fontani2017}; \citealt{favre2018}).

Finally, in order to be able to disentangle the contamination of the protostar from the cloud, we also mapped the northern region around MMS5, MMS2, and CSO3 (see Fig. \ref{fig:omc23}). We chose the area to be mapped considering three main criteria. Firstly, the area has to contain both a part of the filament, with at least one protostar from our sample, and a part outside of the filament to see whether the abundance ratio varies across the map and if so, then how. Secondly, we aimed to target the most quiescent environment possible, excluding regions with sources associated with known outflows or jets \citep{ yu1997, reipurth1999, williams2000, takahashi2008}. These two criteria lead us to select OMC-3, which is more quiescent than OMC-2 from this point of view. Finally, we aimed to target more than one protostar from our sample to understand whether the abundance ratio [CCH]/[CH$_3$OH] varies among the sources and how. Thus, the region centred on CSO3, which includes three of our protostellar targets, was selected as the best area to map.

\subsection{Selected molecules}
In order to identify whether the targeted sources belong to the hot corino, WCCC sources classes, or none of them, we targeted lines from CCH and CH$_3$OH, two molecules that are characteristic of WCCC and hot corinos sources, respectively \citep{ maret2005, sakai2013}. 
As in previous similar works (Section \ref{sec:previous-survesy}), we used their relative abundance ratio, [CCH]/[CH$_3$OH], to assess their chemical nature. Specifically, based on the values measured in the two prototypes of hot corinos and WCCC sources (IRAS16293-2422 and L1527: see Tab. \ref{tab:surveysold}), a [CCH]/[CH$_3$OH] abundance ratio larger than about 2 would roughly identify a WCCC source while a value lower than about 0.5 would testify for a hot corino nature; sources with values in between may be intermediate cases.

We selected two bands, at 3 and 1 mm, of the hyperfine structure lines of CCH, and several CH$_3$OH rotational lines which cover a large range ($\sim$7 to $\sim$100 K) of transition upper level energy, in order to derive column densities and temperatures of CCH and CH$_3$OH.
The list of the targeted lines as well as their spectroscopic parameters is reported in Table.~\ref{tab:lines1}.

\begin{table*}
\small
\centering
\caption{List of targeted lines and their spectroscopic properties. The last two columns indicate whether the line is observed in the single-pointing observations (S) or in the OTF map (M), and the used telescope. The spectroscopic data for CCH are from \citet{padovani2009} through the CDMS (Cologne Database for Molecular Spectroscopy: \citealt{muller2005}) and for CH$_3$OH, from \citet{xu2008} through the JPL molecular spectroscopy database (Jet Propulsion Laboratory: \citealt{pickett1998}) databases }
\label{tab:lines1}
\begin{threeparttable}
\begin{tabular}{|c|clcc|c|c|}
\hline
\multirowthead{2}{Molecule} &\multirowthead{2}{Transition}& Freq. & E$_{\text{up}}$ & A$_{\text{ij}}$ & Observation mode & \multirowthead{2}{Telescope}\\
         &          & [GHz] &   [K]           &  [s$^{-1}$]     & Single-point (S)/Map (M)   &\\
\hline 
\multirow{20}{*}{CH$_{3}$OH}& 5$_{-1}$ - 4$_0$  E& \ \ 84.521&40.4 & 1.97E-06&S and M & IRAM-30m + Nobeyama-45m\\ 
&2$_{-1}$ - 1$_{-1}$  E&\ \ 96.739 & 12.5 & 2.56E-06 & M&IRAM-30m\\
&2$_{0}$ - 1$_0$  A&\ \ 96.741 & 6.9 & 3.41E-06 & M&IRAM-30m\\
&2$_{0}$ - 1$_0$  E&\ \ 96.745 & 20.0 & 3.42E-06 & M&IRAM-30m\\
&2$_{1}$ - 1 $_1$  E&\ \ 96.755 & 28.0 & 2.62E-06 & M&IRAM-30m\\
& 5$_1$ - 4$_1$ A&239.746& 49.1 & 5.66E-05&S&IRAM-30m\\
& 5$_0$ - 4$_0$ E&241.700  &47.9 & 6.04E-05& S&IRAM-30m\\
&5$_{-1}$ - 4$_{-1}$ E&241.767 & 40.4 & 5.81E-05&S&IRAM-30m\\
&5$_0$ - 4$_0$ A&241.791 & 34.8 & 6.05E-05&S&IRAM-30m\\
 & 5$_3$ - 4$_3$ A&241.832\tnote {a} & 84.6& 3.87E-05&S&IRAM-30m\\
  & 5$_3$ - 4$_3$ A&241.833\tnote {a}& 84.6& 3.87E-05&S&IRAM-30m\\
 &5$_2$ - 4$_2$ A   &241.842&72.5&5.11E-05&S&IRAM-30m\\
 & 5$_3$ - 4$_3$ E  &241.843 & 82.5 & 3.88E-05&S&IRAM-30m\\
 &5$_3$ - 4$_3$ E & 241.852 & 97.5 & 3.89E-05&S&IRAM-30m\\
 &5$_1$ - 4$_1$ E&241.879 &55.9 & 5.96E-05&S&IRAM-30m\\
&5$_2$ - 4$_2$ A & 241.887& 72.5& 5.12E-05 &S&IRAM-30m\\
&5$_2$ - 4$_2$ E&241.904\tnote {a} & 57.1 & 5.03E-05&S&IRAM-30m\\
&5$_2$ - 4$_2$ E&241.904\tnote {a}   & 60.7 & 5.09E-05&S&IRAM-30m\\
&5$_1$ - 4$_1$ A &243.915 & 49.7 & 5.97E-05 &S&IRAM-30m\\
&2$_1$ - 1$_0$ E &261.805 & 28.0 & 5.57E-05&S&IRAM-30m\\
\hline
\multirow{15}{*}{CCH} &$N$=1-0, $J$=3/2-1/2, $F$=1-1& \ \ 87.284&4.2&2.60E-07&S and M& IRAM-30m + Nobeyama-45m\\
&$N$=1-0, $J$=3/2-1/2, $F$=2-1& \ \ 87.316&4.2&1.53E-06&S and M& IRAM-30m + Nobeyama-45m\\
&$N$=1-0, $J$=3/2-1/2, $F$=1-0& \ \ 87.328&4.2&1.27E-06&S and M& IRAM-30m + Nobeyama-45m\\
&$N$=1-0, $J$=1/2-1/2, $F$=1-1& \ \ 87.402&4.2&1.27E-06&S and M& IRAM-30m + Nobeyama-45m\\
&$N$=1-0, $J$=1/2-1/2, $F$=0-1& \ \ 87.407&4.2&1.54E-06&S and M& IRAM-30m + Nobeyama-45m\\
&$N$=1-0, $J$=1/2-1/2, $F$=1-0& \ \ 87.446&4.2&2.61E-07&S and M& IRAM-30m + Nobeyama-45m\\
&$N$=3-2, $J$=7/2-5/2, $F$=3-3& 261.978&25.1 & 1.96E-06&S& IRAM-30m\\
 &$N$=3-2, $J$=7/2-5/2, $F$=4-3& 262.004& 25.1 & 5.32E-05&S& IRAM-30m\\
& $N$=3-2, $J$=7/2-5/2, $F$=3-2 & 262.006 & 25.1 & 5.12E-05&S& IRAM-30m\\
 & $N$=3-2, $J$=5/2-3/2, $F$=3-2& 262.064& 25.2& 4.89E-05&S& IRAM-30m\\
 & $N$=3-2, $J$=5/2-3/2, $F$=2-1& 262.067& 25.2 & 4.47E-05&S& IRAM-30m\\
 & $N$=3-2, $J$=5/2-3/2, $F$=2-2 & 262.078& 25.2 & 6.02E-06&S& IRAM-30m\\
& $N$=3-2, $J$=5/2-5/2, $F$=3-3& 262.208& 25.2 & 3.96E-06&S& IRAM-30m\\
& $N$=3-2, $J$=5/2-5/2, $F$=3-2&262.236& 25.2& 4.04E-07&S& IRAM-30m\\
& $N$=3-2, $J$=5/2-5/2, $F$=2-2& 262.250& 25.2& 2.27E-06&S& IRAM-30m\\
\hline
\end{tabular}
\tablefoot{
\tablefoottext{a}{ Lines blended, they are not used in the LTE and LVG analysis.}
}

\end{threeparttable}
\end{table*}

\section{Observations}\label{sec:observations}

We obtained single-pointing observations towards the nine sources of Tab. \ref{tab:sources} with the single-dish telescopes IRAM-30m, targeting the CCH and CH$_3$OH lines in the 3 mm and 1 mm bands, and Nobeyama-45m, for additional 3mm CCH lines. In addition, in order to disentangle the inner protostellar emission from that originating from the more external parental cloud, we obtained a $3'\times3'$ arcmin$^2$ map of the north end of OMC-3 with the IRAM-30m (see Fig. \ref{fig:omc23}) in both CCH and CH$_3$OH lines. In the following, we individually describe  the observations obtained at the two telescopes.

\subsection{IRAM-30m telescope}

Single-pointing observations were carried out on January 13$^{\text{th}}$ and 17$^{\text{th}}$ 2016, and on February 24-25$^{\text{th}}$ and 29$^{\text{th}}$ 2016. The EMIR receiver E2 (1.3mm) was used in order to cover the chosen spectral windows from 239.14 GHz to 246.92 GHz and from 254.82 GHz to 262.6 GHz. The Fourier Transform Spectrometer (FTS), providing a spectral resolution of 195 kHz (0.26 km.s$^{-1}$),  was connected to the receiver. For each source, single-pointing observations in wobbler mode were made, with a wobbler throw of 120$''$. The resulting beam size is 10$''$. Focus and pointing were checked every 1.5 hours and every six hours respectively throughout the observations and with a pointing accuracy of $\leq 2$ arcsec at 1mm and of $\leq 5$ arcsec at 3mm.

Additionally, an on-the-fly (OTF) map of size $3'\times3'$ arcmin$^2$, centred on the source CSO3 was observed on October 22$^{\text{nd}}$ 2018 (see Fig.~\ref{fig:omc23}) with a beam size of 30 $''$. The dump time was 1 s and the sampling interval 10$''$. The map was repeatedly scanned both along the right ascension (R.A.) and the declination (Dec.) directions until the required root mean square (rms) was reached.  We used the EMIR receiver and the FTS units to cover the frequencies from 80.2 GHz to 87.8 GHz and from 95.7 GHz to 103.3 GHz. The spectral resolution obtained is 0.7 km/s.

We used the Continuum and Line Analysis Single-dish Software (CLASS) from the GILDAS package\footnote{\textit{http://www.iram.fr/IRAMFR/GILDAS}.} to reduce the two sets of data. For the single-pointing observations,  a baseline of a first-order polynomial was subtracted for each scan. Then all spectral scans were stitched together to get a final spectrum for each source. Finally, the intensity was converted from antenna temperature ($T_{\text{A}}^{*}$) to main beam temperature ($T_{\text{mb}}$) using the Ruze's formula $B_{eff}=B_0.\text{exp}[-(4\pi\sigma/\lambda)^2]$, the scaling factor B$_0$= 0.863 and the width factor $\sigma = 66 \mu \text{m}$ (from the IRAM website\footnote{\textit{http://www.iram.es/IRAMES/mainWiki/Iram30mEfficiencies}.}). 
For the OTF map, the baseline was corrected before conversion from $T_{\text{A}}^{*}$ to $T_{\text{mb}}$. 

Through an inspection of individual scans, the calibration uncertainties were estimated to be better than 20\% and 15\% for the single-pointing and the OTF observations, respectively. The rms sensitivity for a channel width of 195 kHz is in the range of 9.2-9.6 mK [$T_{\text{A}}^{*}$] for single-pointing and is 20 mK [$T_{\text{A}}^{*}$] for the OTF map. The spectra of the source SIMBA-a showed emission at the offset position thus contaminating the final spectrum. This source has thus been left out from any further analysis concerning the IRAM-30m observations.

\subsection{Nobeyama-45m telescope}
The observations were carried out between the 10$^{\text{th}}$ and the 17$^{\text{th}}$ of January 2016. The T70 receiver was used in order to cover the chosen spectral windows from 72 to 76 GHz and from 84 to 88 GHz in the 3mm band. A bank of 16 SAM45 auto-correlators were used with a spectral resolution of 244 kHz. The resulting beam size is 20 $''$. For each source, position-switch observations were made. Pointing of the telescope was checked by using Orion KL SiO maser emission \citep{snyder1974,wright1983} every 1.5 hours throughout the observations with a pointing accuracy of about 4 arcsec.

After inspection of individual scans, the calibration uncertainties were estimated to be better than 40\%. The main-beam temperature ($T_{\text{mb}}$) was derived using the telescope main beam efficiency, which is obtained by comparing the CCH (87.316 GHz) and H$_2$CO (72.84 GHz) lines  observed towards Orion KL, the reference source.  We used the software JNewstar\footnote{\textit{https://www.nro.nao.ac.jp/\textasciitilde jnewstar/html}} developed by the Nobeyama Radio Observatory to reduce the data. The rms sensitivity acquired for a channel width of 244 kHz is 5 mK [$T_{\text{A}}^{*}$]. 

\section{Results}\label{sec:results}

In all three data sets, the two single-pointing observations with IRAM-30m and Nobeyama-45m plus the map with IRAM-30m, we extracted the spectra of the lines in Tab. \ref{tab:lines1} and fitted the detected ones with Gaussian functions in order to derive the usual parameters, the velocity-integrated intensity, rest velocity, and line full width at half maximum (FWHM).

\subsection{Single-pointing}\label{results_singlepoint}
\begin{figure*}
\centering
\includegraphics[width=17cm]{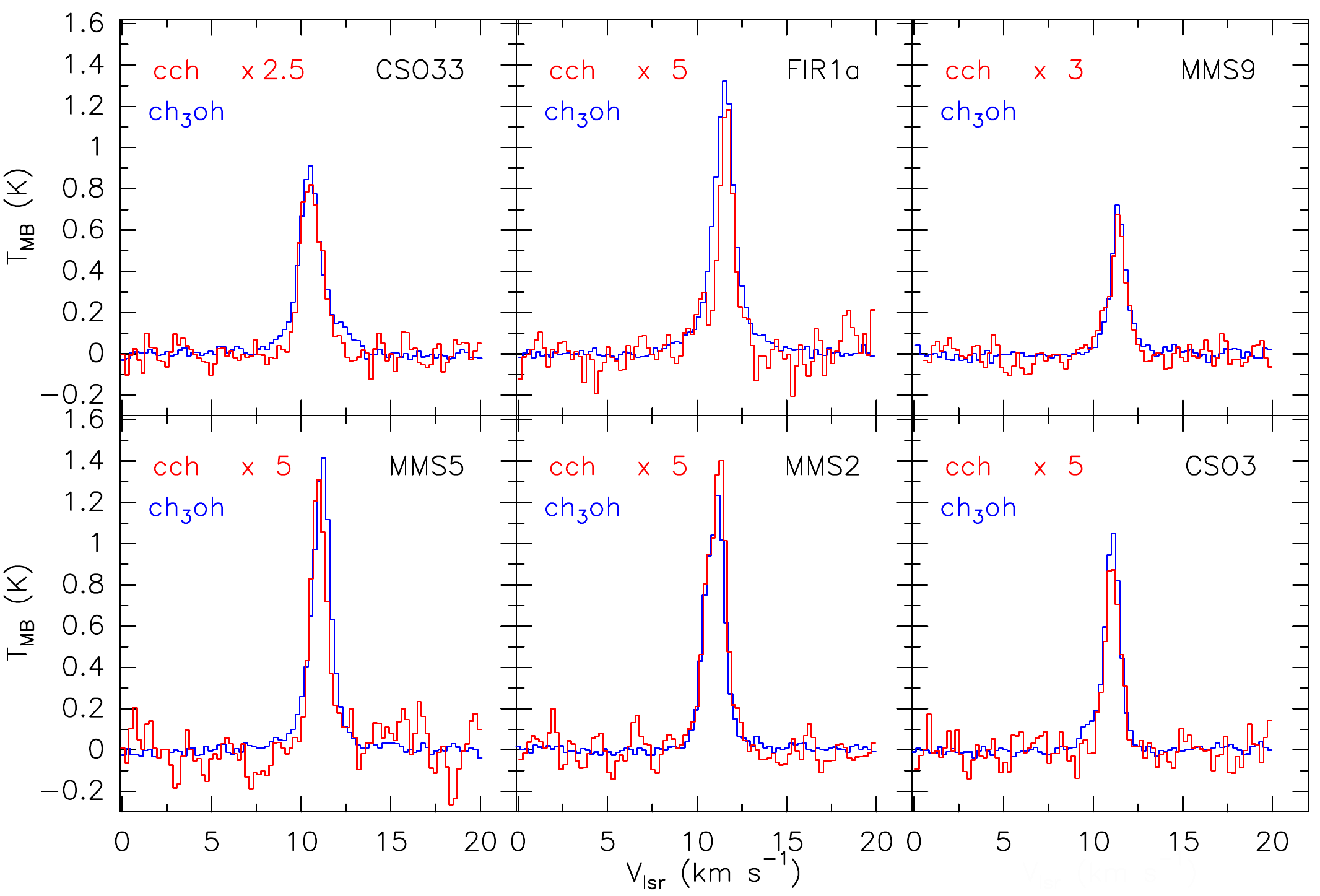}
\caption{Spectra of the CCH ($N$=3,2, $J$=5/2-3/2, $F$=2-2; red) and CH$_3$OH (5$_{0}$-4$_{0}$ A; blue) lines observed with the IRAM-30m telescope towards six sources of the sample that show well-constrained line profiles suitable for the analysis (see Section \ref{results_singlepoint}) and for which 1mm data are available (see Tab. \ref{tab:sources}). For the purposes of visibility,
the spectra of CCH has been increased by various scale factors in order to
match the size of the CH$_3$OH lines.}
\label{fig:sup_cch_ch3oh}
\end{figure*}

\begin{figure}
\centering
\resizebox{\hsize}{!}{\includegraphics{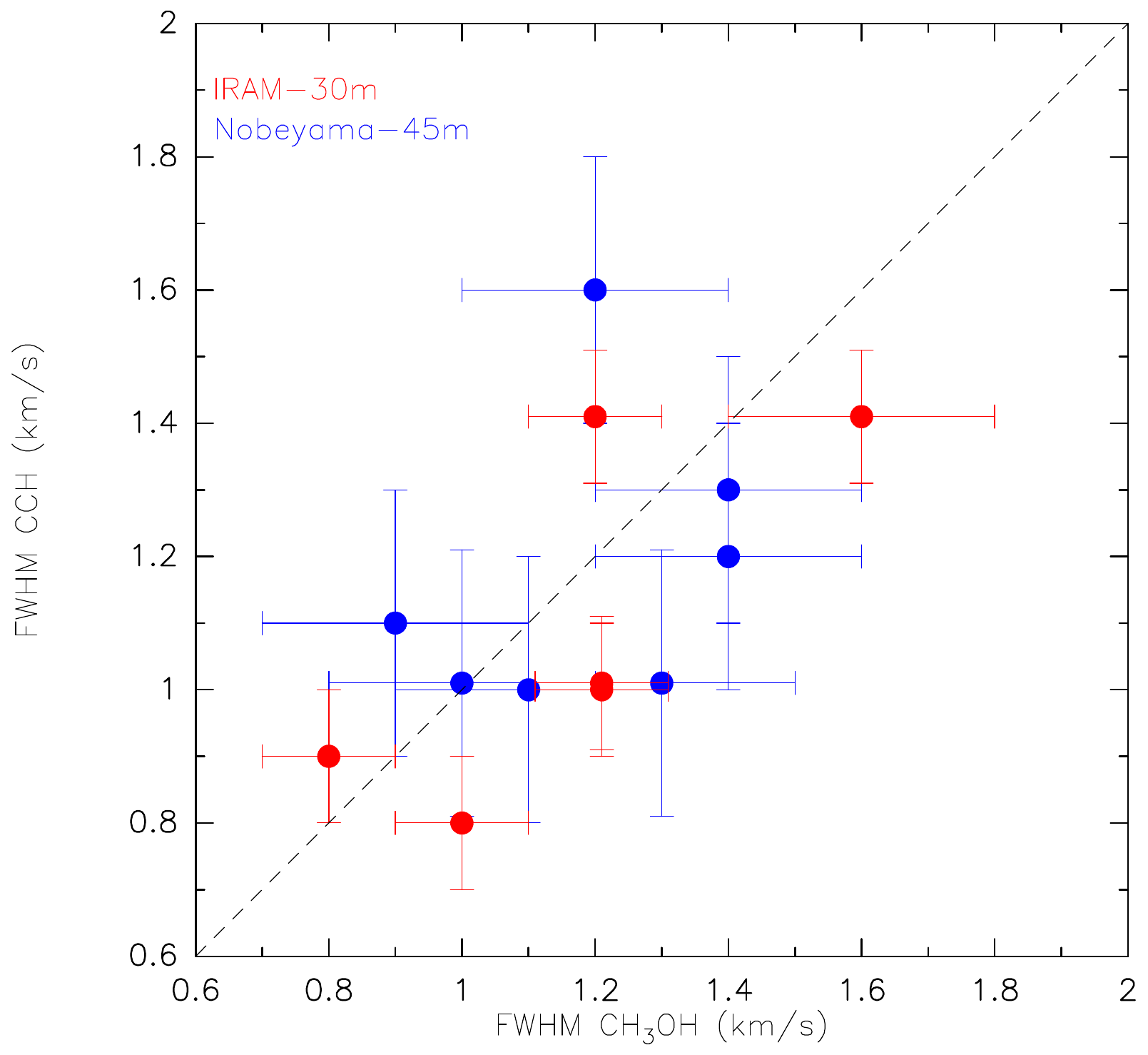}}
\caption{Average widths (FWHM) of the CCH (y-axis) and CH$_3$OH (x-axis) lines observed in each source of Tab. \ref{tab:sources}, using the IRAM-30m (red) and the Nobeyama-45m (blue), respectively. The errors are 1$\sigma$ of the average values. This shows that the two molecules are emitted in a similar and quiescent region (\S\ref{results_singlepoint}). We note that only six sources are reported for the IRAM-30m observations and seven for the Nobeyama-45m: this is because the sources FIR6c and FIR2 have been left out from the analysis in both set of observations, the lines being blended with outflows (See Tab.~\ref{tab:sources}). The source SIMBA-a is contaminated by the wobbler off position only in the IRAM-30m data set. We could thus perform the analysis with Nobeyama-45m (See Tab.~\ref{tab:sources}).}
\label{Fig:comp_FWHM}
\end{figure}

The intensity threshold set for the line detection is 3$\sigma$ on the peak intensity of the line for a channel width of $\approx 0.2$ km/s and of $\approx 0.4$ km/s for the IRAM-30m and Nobeyama-45m data, respectively. 
A summary of all transitions of CCH and of CH$_3$OH that have been detected in the source sample is shown in Tab.~\ref{tab:lines}. Samples of the observed spectra are shown in Fig. \ref{fig:sup_cch_ch3oh}. 
The Local Standard Rest velocities $V_{\text{lsr}}$ derived are in the range 10.5 $-$ 11.6 km/s for both species. We find line widths in the range from 0.8 to 1.6 km/s for CCH and CH$_3$OH both at 1 mm and 3 mm, as shown in Fig.~\ref{Fig:comp_FWHM}. Furthermore, when superimposing the most intense\footnote{The most intense lines of CCH being blended, we chose the next intense one.} lines of CCH and CH$_3$OH  from the IRAM-30m data, as shown in Fig.~\ref{fig:sup_cch_ch3oh}, the similarity of the shape of the lines is evident. The detected lines of each species for each source are reported in Appendix \ref{appdxA}. 

In FIR2, high-velocity components suggestive of outflows have been detected around multiple lines. In particular, CH$_3$OH lines display a prominent high-velocity component that blends with the ambient component at the V$_{\text{lsr}}$ of the source. Disentangling the ambient from the high-velocity component of CH$_{3}$OH lines with the current data set was thus impossible. Similarly, in  FIR6c, important high-velocity emission is present and as a consequence, the lines of CH$_{3}$OH have non-Gaussian shapes. We could not  isolate the component tracing the ambient envelope. These two sources have been left out from further analysis. The $N$=5-4 transition of CH$_3$OH for those two sources is shown in Appendix\ref{appdxA}.

\indent Table~\ref{tab:resultfit-narrow} presents the statistics results from the Gaussian fits. The fit parameters for each molecule and for each source are presented in Appendix \ref{appdxB1}. In most cases, we could fit the lines with a single Gaussian component. The most intense lines of CH$_3$OH and CCH show a broad ($\sim 4$ km/s) component and could be fitted with two Gaussian curves. However, due to the small number of broad components  clearly detected for each source, we do not have enough data to characterise this second component which could be associated with outflowing motions. We will thus focus on the narrow component ($\leq 2$ km/s) of the lines in what follows. \\

\begin{table*}
\small
\centering
\caption{List of all detected and non-detected lines for each source for the single-pointing observations. Y stands for detected (SNR $\geq3$)and N for non-detected. The symbol 'O' indicates that the line is contaminated by the emission of the outflow wing of an adjacent line making the assessment of a detection impossible.}
\label{tab:lines}
\renewcommand{\arraystretch}{1.2}
\begin{tabular}{|c|l|ccccccccc|}
\hline
Molecule & Freq. (GHz)  & CSO33 & SIMBA-a & MMS5 & FIR2$^{\text{a}}$  & FIR6c$^{\text{b}}$  &FIR1a&MMS9&MMS2&CSO3\\
\hline 
\multirow{14}{*}{CH$_{3}$OH}&  \ \ 84.521&Y &Y& Y &Y&O&Y&Y&Y&Y\\
&239.746&Y&...&Y&O&O&Y&Y&Y&N\\
& 241.700  &Y&...&Y&O&O&Y&Y&Y&Y\\
&241.767 & Y&...&Y&O&O&Y&Y&Y&Y\\
&241.791 & Y&...&Y&O&O&Y&Y&Y&Y\\
& 241.832 &N&...&N&O&O&N&N&N&N\\
  & 241.833& N&...&N&O&O&N&N&N&N\\
  &241.842&N &...&N &O & O&N &N &N &N\\
 &241.879 &Y&...&Y&O&O&Y&Y&Y&Y\\
& 241.887&N&...&Y&N&O&N&N&N&N \\
&241.904 & Y&...&Y&O&O&Y&Y&Y&Y\\
&241.904  & Y&...&Y&O&O&Y&Y&Y&Y\\
&243.915 & Y&...&Y&Y&O&Y&Y&Y&N\\
&261.805 & Y&...&Y&Y&O&Y&Y&Y&Y\\
\hline
\multirow{15}{*}{CCH} & \ \ 87.284&Y&Y & Y &O&Y&Y&Y&Y&Y\\
& \ \ 87.316&Y & Y&Y &O&Y&Y&Y&Y&Y\\
& \ \ 87.328&Y & Y&Y &O&Y&Y&Y&Y&Y\\
& \ \ 87.402&Y & Y &Y&O&Y&Y&Y&Y&Y\\
& \ \ 87.407&Y & Y& Y&O&Y&Y&Y&Y&Y\\
& \ \ 87.446&Y & Y &Y&O&Y&Y&Y&Y&Y\\
& 261.978&Y&...&Y&O&Y&Y&Y&Y&Y\\
 & 262.004&Y&...&Y&O&Y&Y&Y&Y&Y\\
&  262.006 &Y&...&Y&O&Y&Y&Y&Y&Y\\
 &  262.064& Y&...&Y&O&Y&Y&Y&Y&Y\\
 &  262.067& Y&...&Y&O&Y&Y&Y&Y&Y\\
 &  262.078&Y&...&Y&O&Y&Y&Y&Y&Y\\
& 262.208&Y&...&Y&Y&Y&Y&Y&Y&Y\\
& 262.236& N&...&Y&Y&N&N&N&N&N\\
&  262.250&Y&...&Y&Y&Y&Y&Y&Y&Y\\
\hline
\end{tabular}

\begin{tablenotes}[para,flushleft]
\footnotesize
\item[a] Source left out of analysis due to prominent outflow features (see Tab.~\ref{tab:sources}) 
\item [b]Source left out of analysis due to non-Gaussian shapes in the line profile  (see Tab.~\ref{tab:sources})
\end{tablenotes}
\end{table*}

\begin{table*}
\centering
\makegapedcells
\setcellgapes{4pt}
\caption{Mean derived properties from Gaussian fits for each source and for each set of observations.}
\label{tab:resultfit-narrow}
\resizebox{\textwidth}{!}{
\begin{tabular}{|c|cc|cc|cc|cc|}
\hline 
\multirowthead{4}{Source}& \multicolumn{2}{c|}{\textit{IRAM}} &\multicolumn{2}{c|}{\textit{Nobeyama}}&\multicolumn{2}{c|}{\textit{IRAM} } &\multicolumn{2}{c|}{\textit{Nobeyama} }\\
\cline{2-9}
 & $ V_{\text{lsr CH}_3\text{OH}}$&$ V_{\text{lsr CCH}}$ & $ V_{\text{lsr CH}_3\text{OH}}$&$ V_{\text{lsr CCH}}$ & $ \text{FWHM}_{\text{CH}_3\text{OH}}$ & $ \text{FWHM}_{\text{CCH}}$ &$ \text{FWHM}_{\text{CH}_3\text{OH}}$ & $ \text{FWHM}_{\text{CCH}}$ \\
& [km.s$^{-1}$]&  [km.s$^{-1}$] & [km.s$^{-1}$]& [km.s$^{-1}$] & [km.s$^{-1}$]&  [km.s$^{-1}$] &  [km.s$^{-1}$]&  [km.s$^{-1}$]  \\
\hline
CSO33 & 10.5 $\pm$ 0.1& 10.5 $\pm$ 0.1&10.7 $\pm$ 0.2&10.6 $\pm$ 0.2&1.6 $\pm$ 0.2&1.4 $\pm$ 0.1&1.4 $\pm$ 0.2&1.3 $\pm$ 0.2\\
FIR1a &11.6 $\pm$ 0.1&11.6 $\pm$ 0.1&11.3 $\pm$ 0.2&11.3 $\pm$ 0.2&1.2 $\pm$ 0.1&1.0 $\pm$ 0.1&1.4 $\pm$ 0.2&1.2 $\pm$ 0.2 \\
MMS9 &11.4  $\pm$ 0.1&11.4 $\pm$ 0.1&11.5 $\pm$ 0.2&11.6 $\pm$ 0.2&1.0 $\pm$ 0.1&0.8 $\pm$ 0.1&1.3 $\pm$ 0.2&1.0 $\pm$ 0.2 \\
MMS5 &11.1  $\pm$ 0.1&11.0 $\pm$ 0.1&11.4 $\pm$ 0.2&11.2 $\pm$ 0.2&1.2 $\pm$ 0.1&1.0 $\pm$ 0.1&1.1 $\pm$ 0.2&1.0 $\pm$ 0.2\\
MMS2 &11.1  $\pm$ 0.1&11.2 $\pm$ 0.1&11.1 $\pm$ 0.2&10.8 $\pm$ 0.2&1.2 $\pm$ 0.1&1.4 $\pm$ 0.1&1.2 $\pm$ 0.2&1.6 $\pm$ 0.2\\
CSO3 &11.1  $\pm$ 0.1&11.1 $\pm$ 0.1&11.2 $\pm$ 0.2&10.9$\pm$ 0.2&0.8 $\pm$ 0.1&0.9 $\pm$ 0.1&1.0 $\pm$ 0.2&1.0 $\pm$ 0.2\\
SIMBA-a & ... & ... &10.5  $\pm$ 0.2 & 10.5 $\pm$ 0.2& ... & ... &0.9 $\pm$ 0.2 &1.1 $\pm$ 0.2 \\
\hline
\end{tabular}
}
\end{table*}

\subsection{Maps}

\begin{figure}
\resizebox{\hsize}{!}{\includegraphics{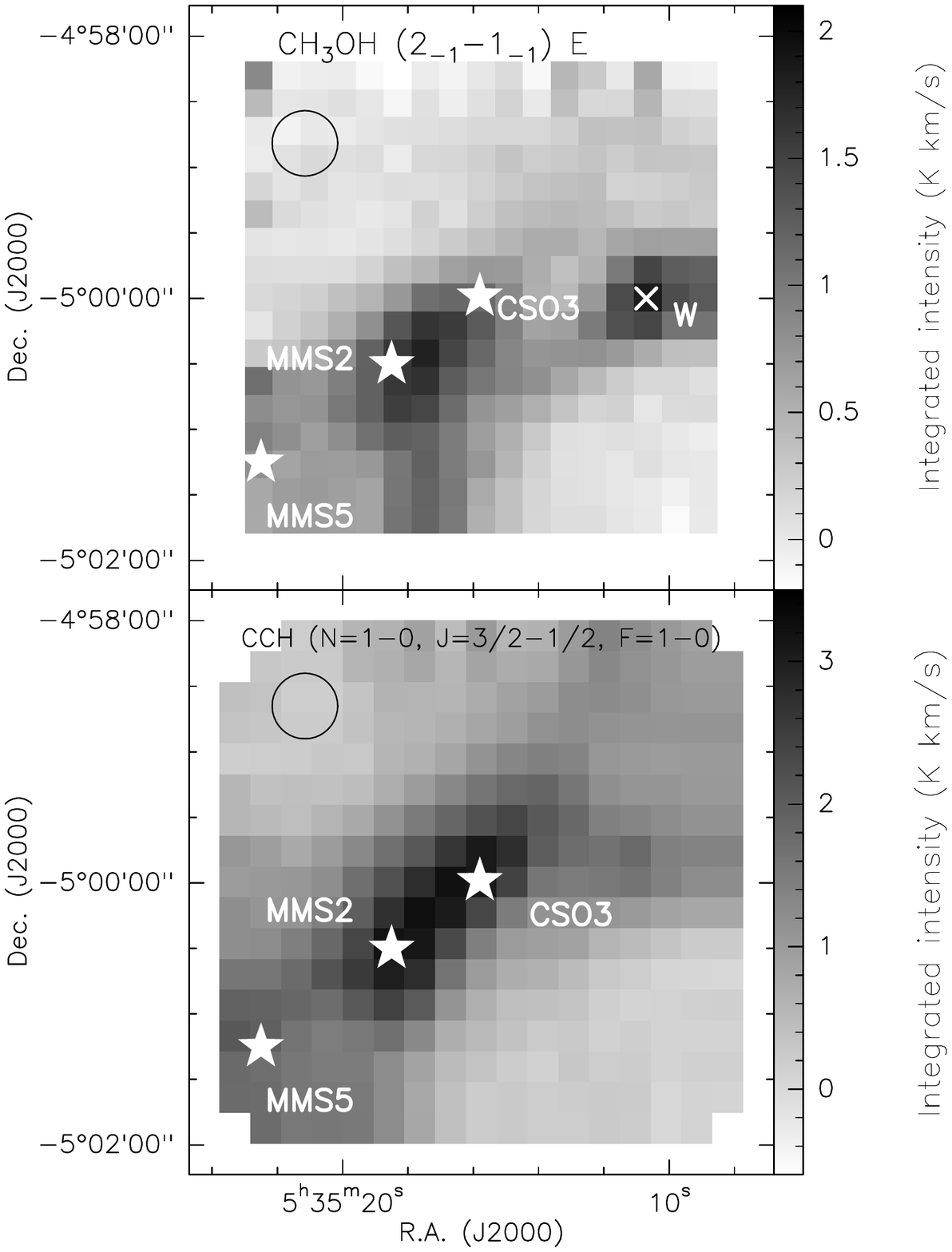}}
\caption{Map of CH$_3$OH(2$_{-1}$ - 1$_{-1}$ E) and CCH ($N$=1-0, $J$=3/2-1/2, $F$=1-0) lines, integrated between 7.9 km/s and 12.8 km/s. The stars represents the sources MMS5, MMS2 and CSO3. The West peak of methanol is represented by a white cross. The filled circle represent the IRAM-30m beam size of 30$''$.}
\label{fig:maps}
\end{figure}

The intensity threshold set for the line detection is 3$\sigma$ on the peak intensity of the line and for a channel width of $\approx 0.7$ km/s. Figure~\ref{fig:maps} shows velocity-integrated maps of the CCH ($N$=1-0, $J$=3/2-1/2, $F$=1-0) and of the CH$_3$OH (2$_{-1}$ - 1$_{-1}$ E) lines. The integration has been done in the velocity range of 7.9 $-$ 12.8 km/s for both lines. The sources MMS5, MMS2, and CSO3 are in the field of the mapped region. In Fig.~\ref{fig:maps}, we can see that the two lines peak at the position near the sources MMS2 and CSO3. In addition, CH$_3$OH shows two other peaks, south and west of the protostars. The emission of CCH is more extended than that of the CH$_3$OH. For clarity, we will refer to the West peak of methanol as the position 'W'. Based on the CH$_3$OH observations, the peak which is located south of the sources can be explained by the presence of outflows that are driven by MMS5 and MMS2 (also seen in CO by \citealt{takahashi2008}). The methanol peak at position 'W' does not seem to correspond with any outflow. The reason of this peak is thus unknown and the we will not discuss any further its origin as it is not the goal of this paper. All the lines presented in Table~\ref{tab:lines1} have been detected. For CCH,the most intense line at 87.316 GHz is detected in 99\% of the map. The next two most intense lines (87.328 and 87.402 GHZ) are detected in 90\% of the map. The line at 87.407 GHz is detected in 85\% of the map. Finally, the two least intense lines (87.284 and 87.446 GHz) are detected in 58 \% of the map. For CH$_3$OH, the most intense lines (96.741 and 96.739 GHz) are present on 65\% of the map on average. Then the lines at 84.521 and 96.445 GHz are present on 30\% of the map. Finally, the least intense line at 96.755 GHz is present on 6\% of the map and its emission is concentrated at the methanol peak at position 'W'. 

Moments 1 and 2 for those same lines are shown in Fig.~\ref{fig:moments}. From the moment maps, we see that the velocities are rather constant throughout the map with a low-velocity region at the location of the second peak of  CH$_3$OH. The line widths of CH$_3$OH are slightly larger (1.5 km/s instead of 1 km/s in the rest of the filament) where the methanol peaks, south-east of the sources. Considering that the sources MM5 and MMS2 drive outflows, this may be the reason of this slight increase.  For CCH, the line width is on average 1 to 1.5 km/s towards the centre and gets larger ($> 2.0$ km/s) to the North West. From the line shapes, we find that the CCH is double peaked in the north of the region, explaining the increase of the line width. A further analysis of the line shape shows that there is a second component located at a rest velocities at 10 km/s in the North west of the filament and 9 km/s where there is the CH$_3$OH peak. This specific line shape is only seen in the CCH and not in the CH$_3$OH lines suggesting an even more external component with mainly CCH in the north of the filament. This could be explained by the fact that CCH is more enhanced by the UV photons than CH$_3$OH. This CCH component would thus probe the outer layers of the cloud, an area that is more exposed to the interstellar field. In the following, we will consider only the component of CCH that is common with the CH$_3$OH, as the second component of CCH does not coincide with the position of the sources and as our goal is solely to place constraints on the region of emission in which both CCH and CH$_3$OH are present.

\section{Physical parameters and [CCH]/[CH$_3$OH] abundance ratios}\label{sec:phys-parameters}

\subsection{Description of the modelling}\label{sec:descr-model}
We used two methods to derive the physical parameters of the emitting gas and the [CCH]/[CH$_3$OH] abundance ratios. We first used an approach based on the Local Thermal Equilibrium (LTE) approximation (as a consistency check) and then we carried out a non-LTE analysis. 

In the LTE approach, we used the usual rotational diagram for the CH$_3$OH lines and the fitting of the hyperfine structure for the CCH lines. Both methods allow us to derive the rotational or excitation temperature and the beam-averaged column density, assuming that the emission is extended, as shown by the maps (Section~\ref{sec:results}). In addition, the rotational diagram approach assumes that the lines are optically thin, an assumption which we verified a posteriori to be correct (see below).

For the non-LTE analysis, we used the Large Velocity Gradient (LVG) model developed by \citet{ceccarelli2003}.
For CH$_3$OH, we used the CH$_3$OH-H$_2$ collisional rates from \citet{flower2010} through the BASECOL database\footnote{\textit{https://basecol.vamdc.eu}} \citep{dubernet2013}. Please note that for the LVG analysis, we used the CH$_3$OH-E species and assumed a ratio CH$_3$OH-E/CH$_3$OH-A equal to 1 to retrieve the total column density of CH$_3$OH and thus calculate the abundance ratio [CCH]/[CH$_3$OH].  For CCH, we used the CCH-H$_2$ collisional rates, from \citet{dagdigian2018}. 
For each source, we ran large grids of models varying the model parameters: the CH$_3$OH-E and CCH column density, $N_{\text{CH$_3$OH-E}}$ and $N_{\text{CCH}}$, from $5\times 10^{11}$ to $1\times 10^{17}$ \pcm \ the kinetic temperature, $T_{\text{kin}}$, from 2 to 200 K; the gas density, n(H$_2$), from $5\times 10^{3}$ to $1\times 10^{10}$ \pcmc. Those ranges of parameters have been chosen according to the values typically found in molecular cores, PDRs, and protostellar envelopes, as we expect the emission to come from those types of environments. The source size was fixed and set as extended. We ran beforehand the LVG model for the source MMS2, present in the OTF map, with the source size set as a free parameter. The result gave a source size significantly larger than the beam size ($\geq 150''$), confirming the extended characteristic of the source. We thus fixed the source size as extended (the filling factor is thus equal to 1) for the LVG analysis.  Finally, the line width was taken equal to the measured FWHM for each source (Tab. \ref{tab:resultfit-narrow}).

In the following, we present the non-LTE analysis results of both the single pointing observations towards six out of the nine target sources (Tab. \ref{tab:sources}) and the map of the northern cloud (Fig. \ref{fig:maps}). The results of the LTE analysis for both single pointing observations and for the OTF map are presented in Appendix \ref{appdxC}.


\begin{figure*}
\includegraphics[width=17cm]{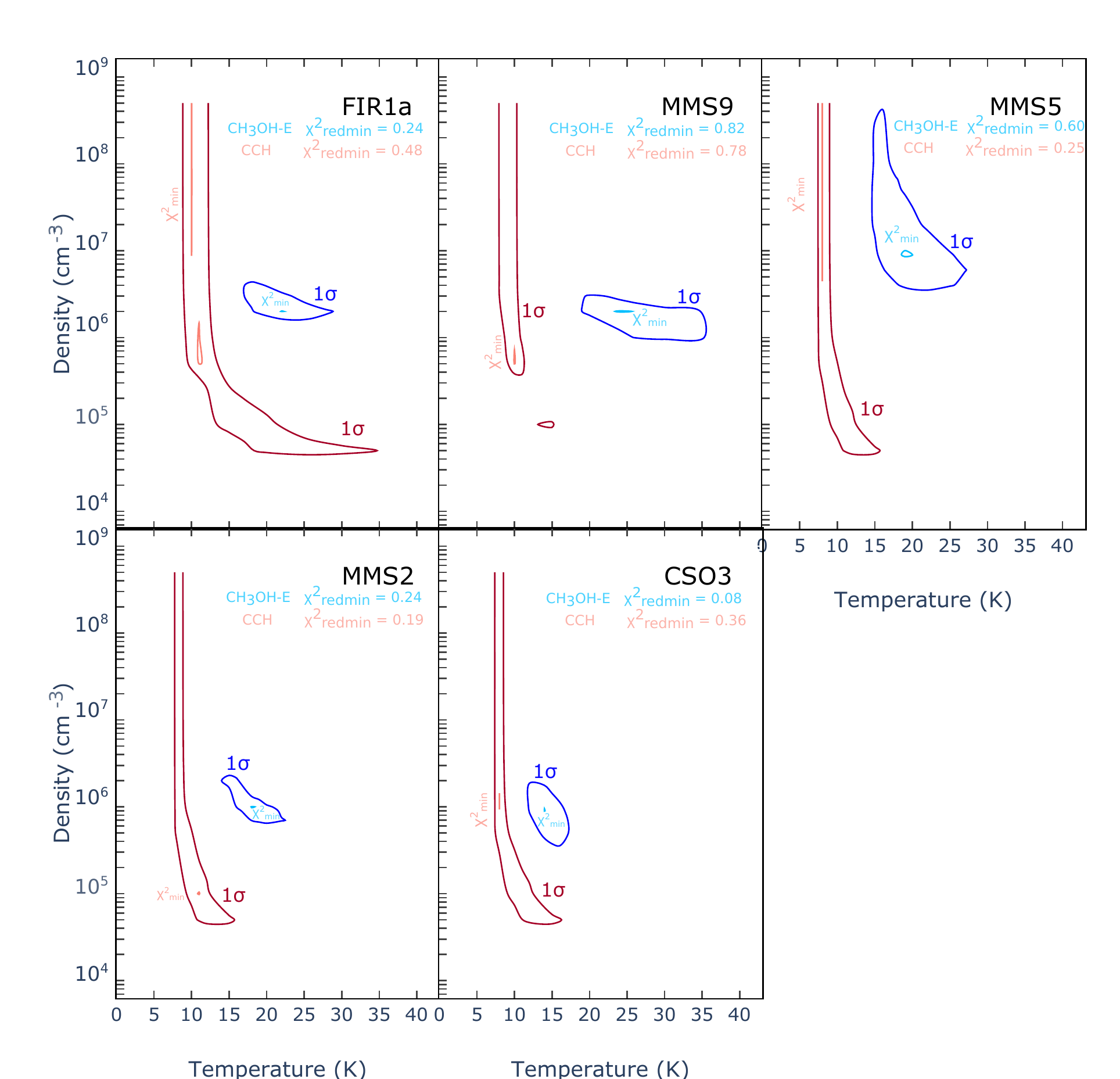}
\caption{Density versus temperature contour plot for all the sources except CSO33 (see Fig.~\ref{fig:lvg_ex}). The minimum $\chi ^2$  and the $1\sigma$ contour plot are obtained for a minimum value of the $\chi^2$ in the column density parameter. The values of the reduced $\chi^2_{\text{redmin}}$ are quoted in the upper corner of each panel.}
\label{fig:lvg_all}
\end{figure*}
\begin{figure}
\centering
\resizebox{\hsize}{!}{\includegraphics{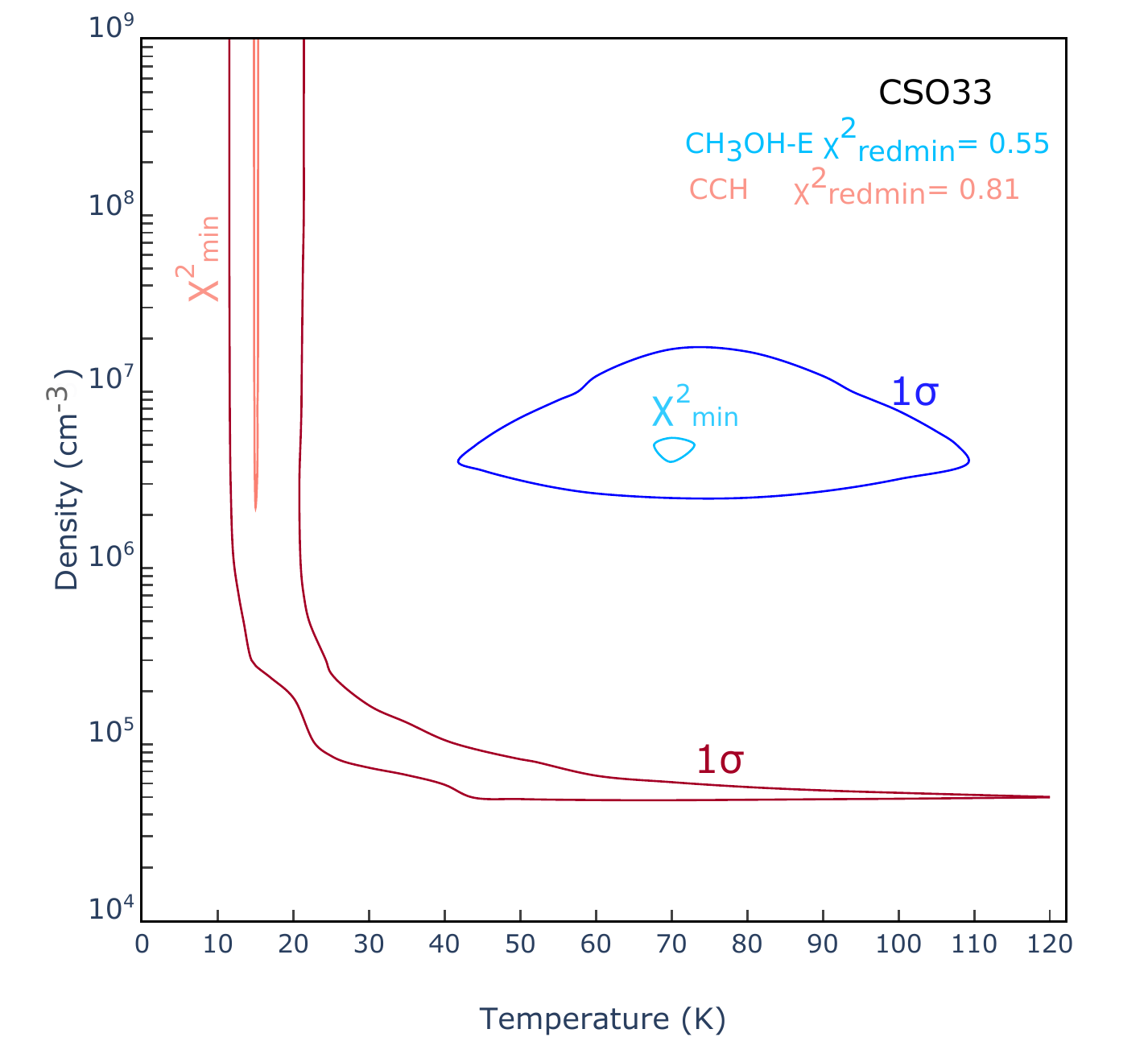}}
\caption{Density versus temperature contour plot for the source CSO33. The source CSO33 represents an extreme case, with the largest temperature of the analysed sample. This is probably because CSO33 is the closest to the Trapezium OB stars cluster. The minimum $\chi ^2$  and the $1\sigma$ contour plot are obtained for a minimum value of the $\chi^2$ in the column density parameter. The values of the reduced $\chi^2_{\text{redmin}}$ are quoted in the upper corner of the panel.}
\label{fig:lvg_ex}
\end{figure}

\subsection{Physical parameters and abundance ratio towards the sources}
We analysed the line emission towards six out of the nine sources of Tab. \ref{tab:sources}. In SIMBA-a, the number of detected lines is too small to obtain meaningful results, while in FIR2 and FIR6c, the line emission is dominated by the outflow.

For each source, the best fit values were found by comparing the LVG model predictions with the observations as follows: for each column density, the minimum $\chi^2$ was found with respect to the density-temperature parameter space, and then the best fit values of density and temperature were derived. The error on each fitted parameter was defined by the 1$\sigma$ distribution. The results of the analysis are summarised in Tab. \ref{tab:lvg_results} and the density-temperature $\chi^2$ contour plots obtained for each source are shown in Figs. \ref{fig:lvg_all} and \ref{fig:lvg_ex} (source CSO33).

The following results are immediately evident, when one excludes CSO33, the source of the sample closest to the Trapezium OB stars cluster:
\begin{enumerate}[label=\arabic*]
    \item All sources have similar gas density and temperature.
    \item While the density and temperature of the gas emitting the methanol lines are well constrained, the density and temperature values are degenerate for the gas emitting CCH. This is because the CCH lines cover a relatively smaller transition upper level energy range with respect to the methanol lines.
    \item The gas emitting methanol and CCH do not have the same density and temperature. Either CCH originates in a less dense or a colder gas, or both, than the gas emitting methanol. The case of MMS9 would suggest that CCH is emitted in a colder gas than methanol.
    \item The gas emitting methanol lines has a temperature that is relatively similar, between 12 and 30 K (considering the errors), and a density also rather similar among the different sources, $(1-3)\times10^6$ \pcmc. These low temperatures are incompatible with the hypothesis that methanol lines originate in a hot corino like region, as we will discuss in more detail in Section~\ref{sec:discussio}.  
    \item In the same vein, the relatively low temperatures, $\leq30$ K, of the gas emitting CCH are compatible with an origin from the cold envelopes of the sources or their parental molecular cloud. 
    \item What is particularly relevant to the goal of this work is that the [CCH]/[CH$_3$OH] abundance ratio is similar in all sources, within the errors, and equal to about 6 (with a range from 1.5 to 13.5 considering the errors).
\end{enumerate}
\begin{table*}
\centering
\makegapedcells
\setcellgapes{2pt}
\caption{Kinetic temperature (T$_{\text{kin}}$),  total column densities (N$_{\text{tot}}$) and abundance ratio [CCH]/[CH$_3$OH] derived from CH$_3$OH-E and CCH.  }
\label{tab:lvg_results}
\resizebox{\textwidth}{!}{
\begin{tabular}{|c|ccc|ccc|c|}
\hline
\multirowthead{2}{Source} & $T_{\text{kin}}$(CH$_3$OH) & $N_{\text{CH}_3\text{OH-E}}$&n(H$_2$)& $T_{\text{kin}}$(CCH) & $N_{\text{CCH}}$&n(H$_2$)&\multirowthead{2}{[CCH]/[CH$_3$OH]} \\
&[K]&[$\times 10^{13}$ cm$^{-2}$]&[$\times 10^{6}$ \pcmc]&[K]&[$\times 10^{14}$ cm$^{-2}$]&[$\times 10^{4}$ \pcmc]& \\
\hline
CSO33&70$^{+40}_{-28}$&2 $^{+3}_{-0.5}$  &5 $^{+7}_{-2}$&15$^{+105}_{-4}$  &3$^{+1}_{-1.5}$  &$>5$&7.5$^{+6}_{-4}$     \\
FIR1a&22 $^{+6}_{-6}$&3 $^{+1}_{-1}$&2$^{+2}_{-0.5}$& 11$^{+24}_{-2}$  &2$^{+1}_{-1}$ &$>4$ &3.5$^{+2}_{-2}$   \\
MMS9&24$^{+11}_{-6}$ &1.5$^{+0.5}_{-0.5}$ &2$^{+1}_{-1}$&  10$^{+5}_{-2}$ & 2$^{+1}_{-1}$ &$>8$&6.5$^{+3.5}_{-3.5}$     \\
MMS5&19 $^{+9}_{-5}$ &3$^{+1.5}_{-1}$  &9$^{+400}_{-6}$& 8$^{+8}_{-1}$  &4$^{+3}_{-2}$  &$>4$&6.5$^{+5}_{-3.5}$     \\
MMS2& 19 $^{+4}_{-4}$  &4$^{+2}_{-1.5}$   &1$^{+1.5}_{-0.3}$& 11$^{+5}_{-3}$  &6$^{+4}_{-3.5}$ &$>4$&7.5$^{+5}_{-3}$    \\
CSO3&14$^{+3}_{-2}$& 3$^{+1.5}_{-1}$  &1$^{+1}_{-0.6}$& 8$^{+8}_{-1}$ & 3$^{+2.5}_{-1.5}$  &$>4$&5$^{+4}_{-3}$    \\
\hline
\end{tabular}
}
\end{table*}

\subsection{Temperature and density of the extended gas}

We also ran the LVG model for the map. However, considering that for CCH we have only the $N$=1-0 transition, it was not possible to do an LVG analysis with this species. For CH$_3$OH-E, we took into account positions with at least three lines to perform the LVG analysis. As for the single-pointing, the gas temperature and gas density derived for each position are based on the best fit for the column density. 

\begin{figure}
 \centering
\resizebox{\hsize}{!}{\includegraphics{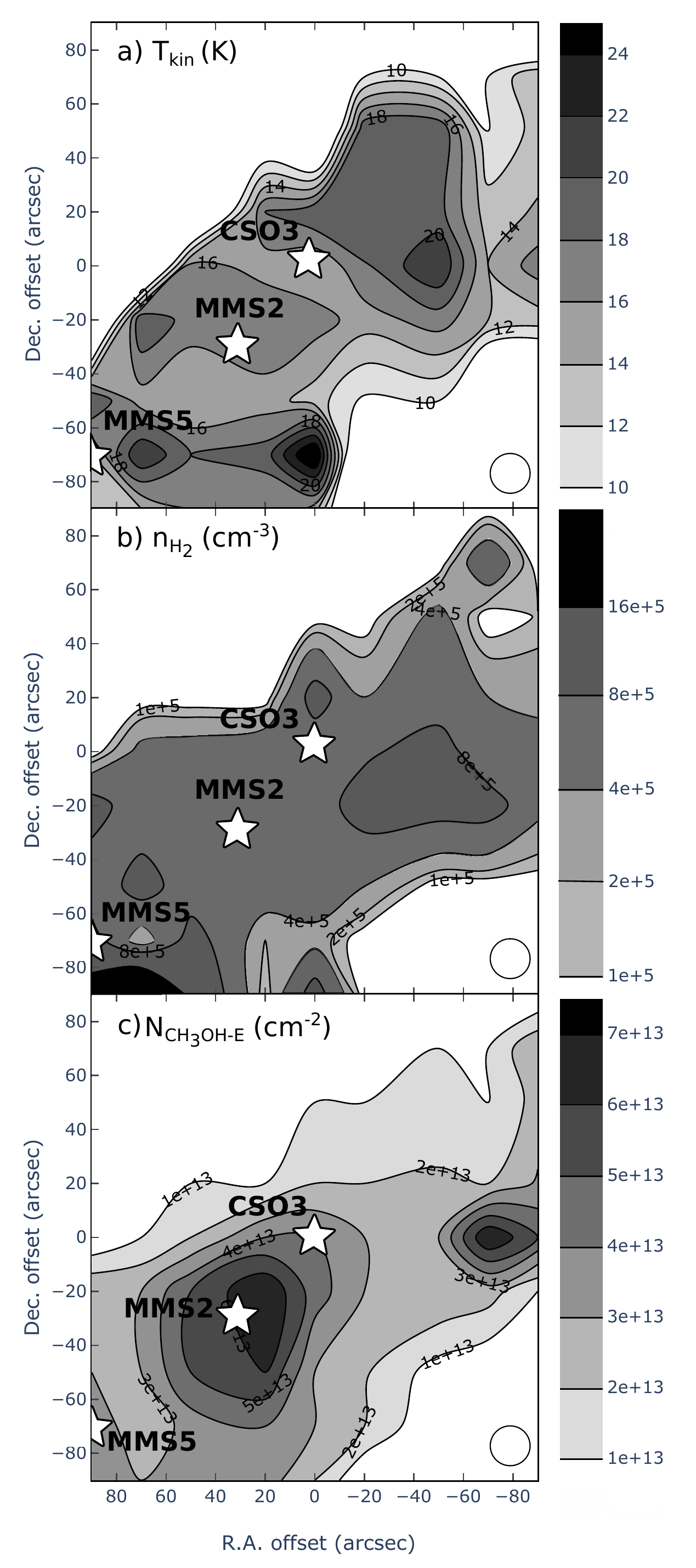}}
\caption{Map of the CH$_3$OH-E column density (bottom panel), temperature (top panel), and density (central panel) of the gas emitting the methanol lines in the northern cloud, which is shown in Fig. \ref{fig:omc23}, as derived by the non-LTE analysis described in Section~\ref{sec:map-aburatio}.  The size of the beam (30$''$) is shown by a white filled circle at the bottom right of each map, the position of the protostars are represented by white stars. We note that the average error bar in the derived temperature is 5--10 K; in the density, it is a factor of two and in the CH$_3$OH-E column density, it is $3\times10^{13}$ cm$^{-3}$. }
\label{fig:maps_lvg}
\end{figure}

The results of the LVG analysis of CH$_3$OH-E are shown in Fig.~\ref{fig:maps_lvg}. 
Within the error bars (about 5--10 K), the temperature of the gas emitting the methanol is rather constant across the cloud, around 15--20 K, with a slightly lower temperature, between 12--10 K, at the border of the cloud.
Again within the error bars (a factor two), the H$_2$ density is also relatively constant across the cloud, with the bulk of the emission originating from a gas at density $(5-9)\times10^6$ \pcmc \ and possibly a slightly less dense halo surrounding it. We verified the non-variation of the properties of the CH$_3$OH emitting gas throughout the filament by deriving the intensity ratio of two optically thin lines of methanol (see Appendix \ref{appdxB2}).
When considering the error bars, both the temperature and the density are very similar to those derived towards the three sources in the field: CSO3, MMS2, and MMS5. In other words, there is no clear evidence that the detected methanol emission originates in the three sources rather than in the parent cloud gas belonging to the cloud. \\
\indent Finally, also the methanol column density is, within a factor of 3, constant across the cloud and no obvious increase is seen in correspondence with CSO3, and only by a factor of 5 towards MMS2 and MMS5 (which lies at the border of the map, so the result has to be taken with caution). 

\begin{figure}
\centering
\resizebox{\hsize}{!}{\includegraphics{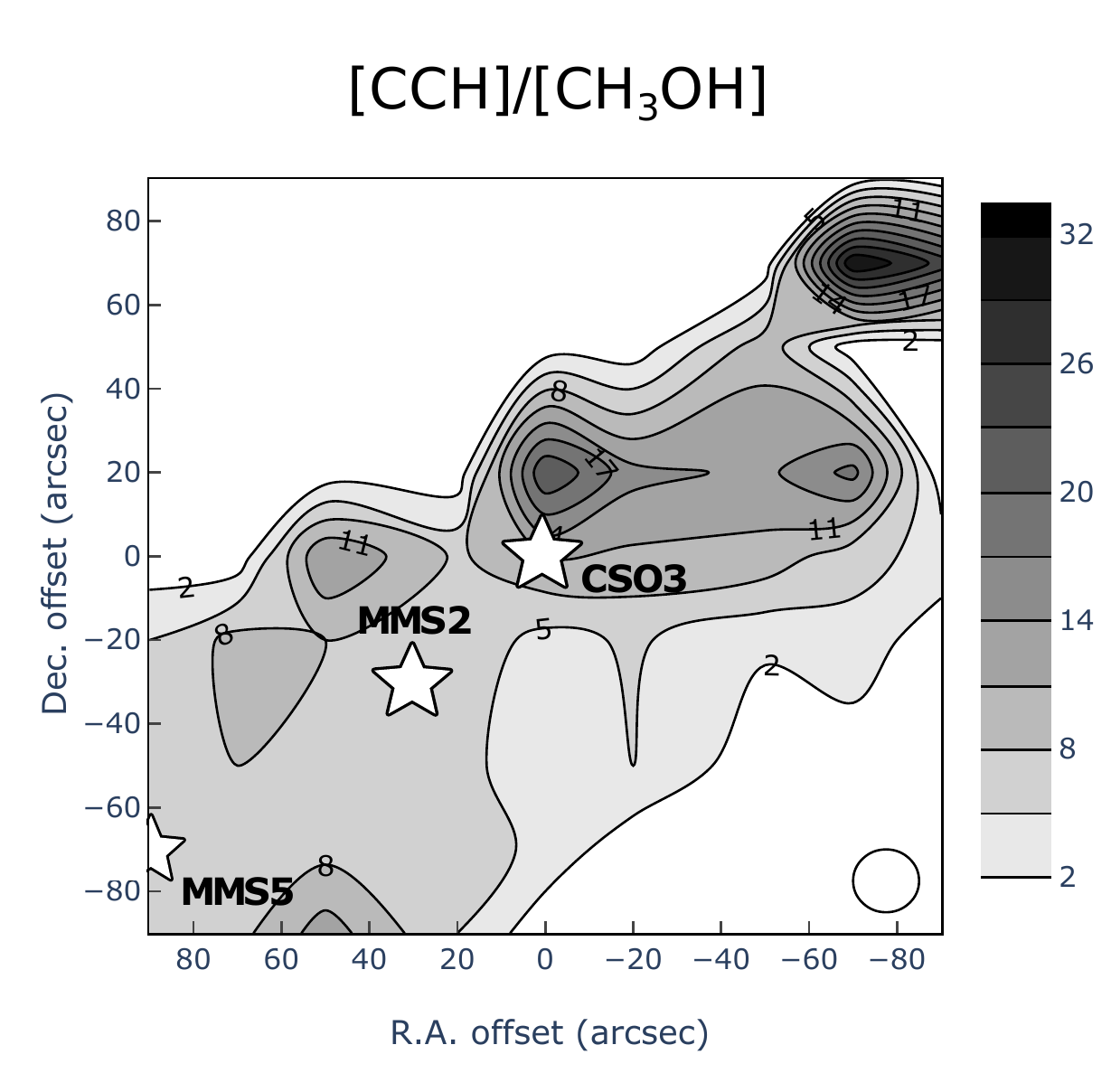}}
\caption{Map of the [CCH]/[CH$_3$OH] abundance ratio as derived via the LTE analysis described in Section~\ref{sec:map-aburatio}. The sources MM5, MMS2, and CSO3 are represented by white stars. The beam size is the white circle at the bottom right of the figure. We note that the error bar in the derived [CCH]/[CH$_3$OH] abundance ratio is about 4 (see Tab. \ref{tab:lte_map} of Appendix \ref{appdxC}).}
\label{fig:LTE_ratio-map}
\end{figure}

 \subsection{[CCH]/[CH$_3$OH] abundance ratio across the northern cloud}\label{sec:map-aburatio}

As we stated in the previous section, in the map of the northern cloud, we could not derive the [CCH]/[CH$_3$OH] abundance ratio via a non-LTE analysis because of the limited number of CCH lines observed. We therefore used the LTE approach for deriving the CCH and methanol column density in each pixel, which is sampled every half beam. In order to derive the abundance ratio, we assume that CCH and CH$_3$OH are emitted in the same region. 
In total, 81 positions throughout the map were extracted and analysed. 
Of those, we took into account positions with at least three lines for both CCH and CH$_3$OH. For CCH we used the hyperfine line structure whereas for CH$_3$OH we used the rotational diagram method, as described in \S~\ref{sec:descr-model}. The results of the LTE analysis are reported in Tab. \ref{tab:lte_map} of Appendix \ref{appdxC}.

The derived total column densities are in the range $(2-11)\times 10^{14}$ cm $^{-2}$ for CCH and in the range $(2- 16)\times 10^{13}$ cm $^{-2}$ for CH$_3$OH, which is in quite good agreement with the values derived by the non-LTE analysis. 
The map of the [CCH]/[CH$_3$OH] abundance ratio is shown in Fig.~\ref{fig:LTE_ratio-map}.
It shows an overall gradient from north to south, with the northern part of the cloud more enriched in CCH with respect to CH$_3$OH than the southern part. On the other hand, we do not see any significant variation of the abundance ratio in correspondence of the positions of the protostars in the field. In other words, the [CCH]/[CH$_3$OH] abundance ratio is more associated with the parent cloud gas in the cloud than with the protostars themselves.

\section{Discussion}\label{sec:discussio}

\subsection{The ambiguous origin of the CCH and CH$_3$OH line emission towards the targeted sources}\label{sec:disc-source-emi}

The first goal of this work is to identify the nature of the surveyed sources, whether they are hot corinos or WCCC objects (Section ~\ref{sec:intro}). As done in previous surveys (Section ~\ref{sec:previous-survesy}), we observed several CCH and CH$_3$OH lines in order to derive their relative abundance ratio and, consequently, to assess the sources chemical nature: it is a hot corino if [CCH]/[CH$_3$OH] is less than about 0.5 and WCCC object if larger than about 2.
Based on this definition and the derived [CCH]/[CH$_3$OH] abundance ratios ($\sim$6: Tab. \ref{tab:lvg_results}), we could argue that all the sources in OMC-2/3 are WCCC objects.
However, before taking this conclusion as final, we have to be sure that the emission towards the sources is due to the sources themselves and that it is not polluted, nor dominated, by the emission from the cloud to which the sources belong -which is the second goal of this study.
We therefore start the discussion on this point first.
\begin{figure*}
\includegraphics[width=17cm]{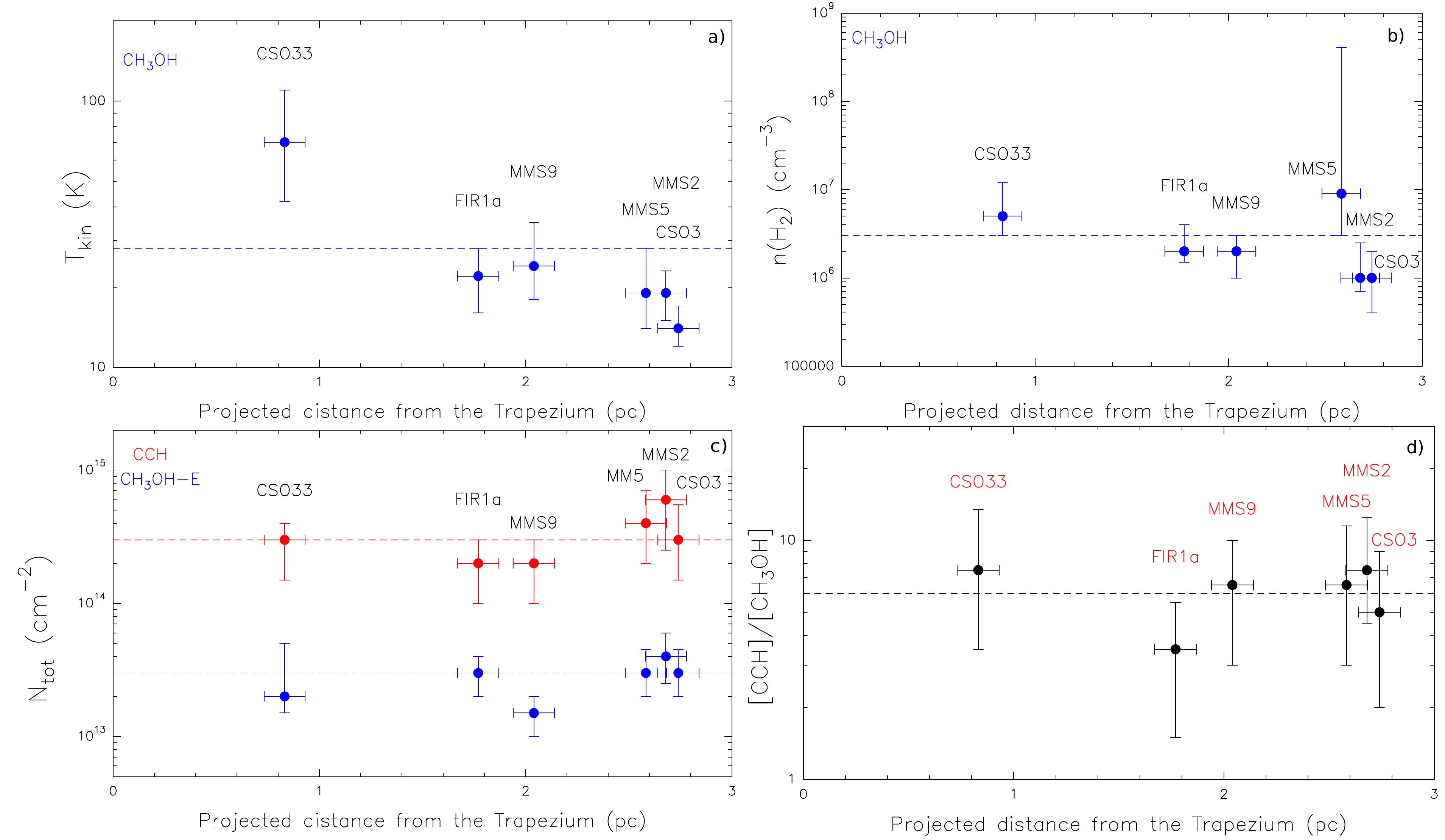}
\caption{Results of the LVG analysis for each source. a) kinetic temperatures (T$_{\text{kin}}$) of CH$_{3}$OH as a function of the distance to the Trapezium; b) Column densities (N$_{\text{tot}}$) of CCH and of CH$_{3}$OH-E as a function of the distance to the Trapezium; c) Derived gas density (n$_{\text{H2}}$) for CH$_3$OH and CCH; d) Abundance ratio [CCH]/[CH$_3$OH-E] as a function of the distance from the Trapezium.  The  red- and blue- dashed lines are the derived averaged values for each parameters for CCH  and CH$_{3}$OH-E respectively. The error bars are at 1 $\sigma$.  }
\label{fig:LVG_results-sources}
\end{figure*}

We begin by summarising the results of our analysis of the observations towards the sources:
\begin{enumerate}[label=\arabic*]
    \item The CCH and CH$_3$OH line widths are narrow (0.8--1.6 km/s: Fig. \ref{fig:sup_cch_ch3oh}) and their spectral shapes (Fig. \ref{Fig:comp_FWHM}) are very similar between them and among the sources: this first result suggests that the lines are emitted approximately in the same region and that this region is rather quiescent.
    \item The temperature of the gas emitting methanol is always lower than 30 K (with the exception of CSO33; see text above: Fig. \ref{fig:lvg_all} and Tab. \ref{tab:lvg_results}), which certainly does not support the hypothesis of an origin in the sources' hot corino. However, the methanol could be associated with the cold envelope of the sources (e.g. \citealt{maret2005}; \citealt{vastel2014}) or the parental cloud. 
    \item The CCH and CH$_3$OH lines probe a gas at likely different temperatures: The CCH lines trace a gas at about 10 K while CH$_3$OH traces a slightly warmer gas, at about 20 K (Tab. \ref{tab:lvg_results}). This could be consistent with the emission arising from the two molecules in the source envelopes, where methanol originates slightly deeper in the envelope (larger density and higher temperature) than CCH. However, the uncertainty on the CCH density-temperature values (Fig. \ref{fig:lvg_all}, \ref{fig:lvg_ex}) also makes it possible that the two molecules lie in the cloud and, specifically, in its surrounding (chemically stratified) PDR.
    \item The CCH and CH$_3$OH-E derived column densities are rather similar among the sources: either the envelopes have all the same characteristics or the emission is associated with the cloud  or its PDR.
    \item To give an idea of the behaviour of the targeted sources over the full OMC-2/3 filament, in Fig. \ref{fig:LVG_results-sources} we show the gas temperature and density, CCH and CH$_3$OH-E column densities, and the [CCH]/[CH$_3$OH] abundance ratio derived towards each source from the non-LTE analysis as a function of the source projected distance from the Trapezium OB cluster. We do not see any variation of any of those quantities along the filament. The only exception is CSO33, which stands out from the other sources as it has a larger gas temperature; however, the difference is not significant and it is probably due to the fact that this source is the closest to the Trapezium OB cluster. This uniformity of the various derived quantities towards the sources regardless of their position in the filament favours the hypothesis that the emission is dominated by the cloud and/or PDR rather than the sources' envelopes.
\end{enumerate}
In summary, based on the single-dish observations, we conclude that the CCH and CH$_3$OH lines are either associated with the cold envelopes of the sources or the cloud to which they belong to or the PDR encircling the cloud, with hints that the first hypothesis for the source envelope origin is likely to be incorrect. 

\begin{figure*}[!ht]
\centering
\includegraphics[width=17cm]{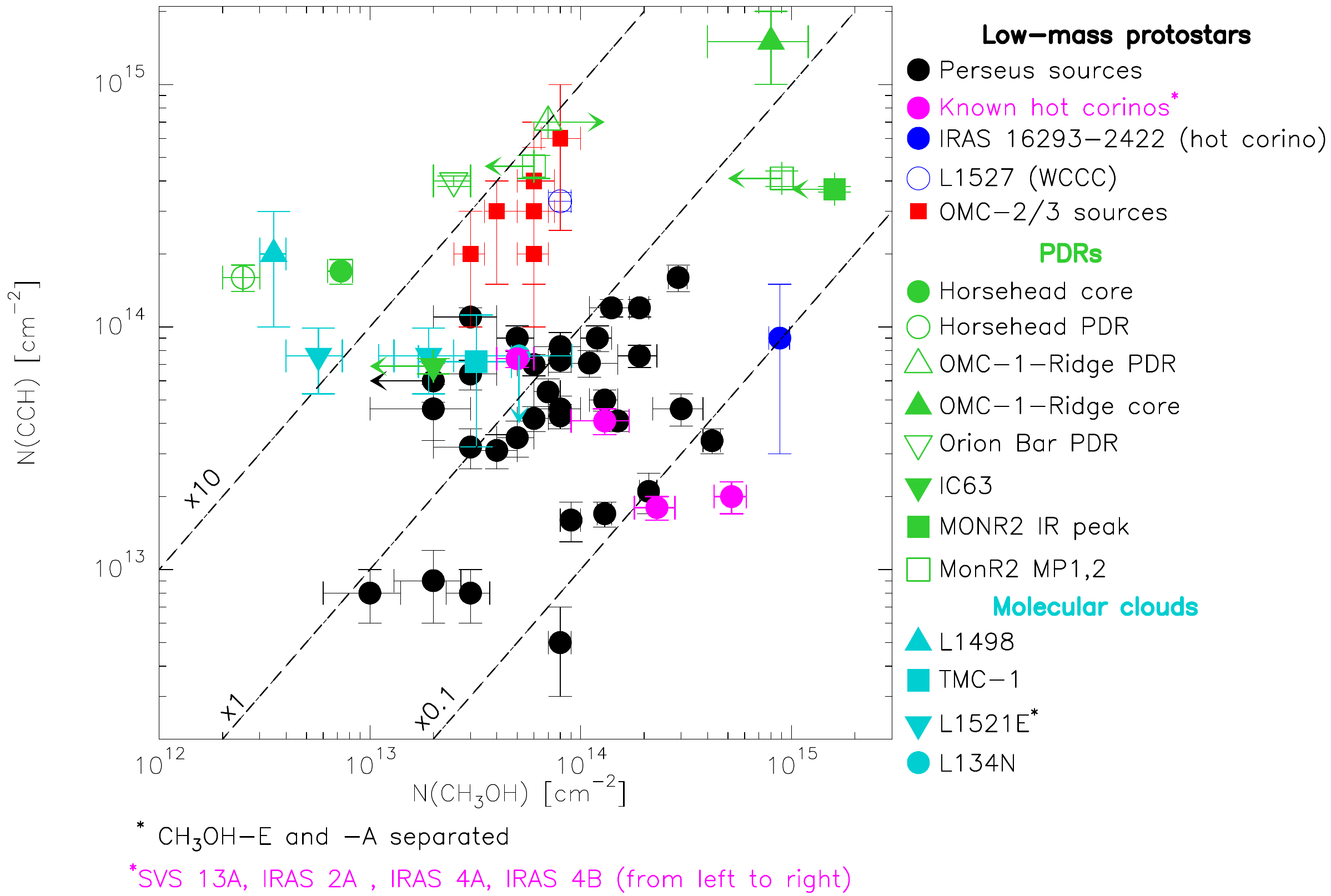}
\caption{Compilation of the CCH (y-axis) and CH$_3$OH (x-axis) total column densities, derived with single-dish observations towards WCCC objects, hot corinos, molecular clouds, and PDRs. The points refer to measurement towards: the WCCC object prototype L1527 \citep[blue open circle:][]{sakai2008b}; the hot corino prototype IRAS 16293-2422 \citep[filled blue circle:][]{vandishoeck1995}; the sources in the Perseus cloud \citep[black:][]{higuchi2018}; the known hot corinos among the Perseus sources \citep[filled pink circle:][]{higuchi2018}, the sources in OMC-2/3 (red: this work); various PDRs (green): Horsehead nebula \citep{teyssier2004,guzman2013}, OMC-1 Ridge PDR and core \citep{ung1997}, Orion Bar  \citep{cuadrado2015,cuadrado2017}, IC63 \citep{jansen1994,teyssier2004}, and Mon R2 \citep{ginard2012}); various pre-stellar cores (light blue): L1498 (\citealt{padovani2009}, \citealt{dapra2017}), TMC-1 (\citealt{friberg88}, \citealt{pratap97}, \citealt{soma2015}), L1521E (\citealt{nagy2019}) and L134N (\citealt{dickens2000}) as representatives of cold molecular clouds. Upper limits are represented by arrows.}
\label{f:fig_hig}
\end{figure*}
%
\subsection{The emission from the cloud and its surrounding PDR}\label{sec:disc-map-emi}
The analysis of the CCH and CH$_3$OH line maps of the northern portion of OMC-3/3 filament (Section~\ref{sec:map-aburatio}) allow us to disentangle the ambiguity described above. We start by summarising the results from the OTF map observations analysis.
\begin{enumerate}[label=\arabic*]
    \item The maps of the gas temperature and density, derived by our non-LTE analysis of the CH$_3$OH-E lines and shown in Fig. \ref{fig:maps_lvg}, clearly indicate that there is no increase (or decrease) of these physical parameters in coincidence with the three sources in the field. The density map shows a sharp increase of the cloud density at its borders and then a ridge with a rather constant density at $(4-8)\times10^5$ cm$^{-3}$. The temperature map also shows a sharp increase at the cloud border and then a rather constant value, 14--16 K, across the ridge.
    \item The CH$_3$OH-E column density shows a gradient which, however, seems associated, rather, with the penetration into the cloud rather than with the sources themselves (Fig.\ref{fig:maps_lvg}): the CH$_3$OH-E column density increases going inwards the central regions of the filament, the ones more shielded by the external photons illumination. There is a marginal increase of the CH$_3$OH-E column density in correspondence of MMS2, from 2$\times 10^{13}$ cm$^{-2}$ to 6$\times 10^{13}$ cm$^{-2}$, namely, about a factor of three with respect to the value in the ridge.
    \item The map of the [CCH]/[CH$_3$OH] abundance ratio  (Fig. \ref{fig:LTE_ratio-map}) does not show any variation caused by the presence of the sources: the [CCH]/[CH$_3$OH] abundance ratio is relatively constant across the region with a possible gradient in the north-Eest to south-west direction, which could be due to the UV illumination from the nearby HII region NGC 1977 (located at $\leq 10''$ from CSO3, so $\leq 2$ pc), located to the North-East of the map.
\end{enumerate}
 When considering the map observations and the derived values (gas temperature and density, CCH and CH$_3$OH-E column densities, and the [CCH]/[CH$_3$OH] abundance ratio) of all the sources along the filament OMC-2/3 (Fig. \ref{fig:LVG_results-sources}), it seems inevitable to conclude that the observed CCH  and CH$_{3}$OH line emission is dominated by the cloud in which the sources are embedded or the PDR that surrounds it.


\begin{table*}
\centering
\caption{Comparison of the results derived from the LTE analysis of the \citet{higuchi2018} survey with the results derived in the present work (both LTE and non-LTE LVG results.} 
\label{tab:surveys}
\makegapedcells
\setcellgapes{2pt}
\begin{tabular}{|c|c|c|c|c|c|}
\hline
\multirowthead{2}{Paper}& T$_{\text{CCH}}$&N$_{\text{CCH}}$&T$_{\text{CH}_3\text{OH}}$&N$_{\text{CH}_3\text{OH-E}}$&\multirowthead{2}{[CCH]/[CH$_3$OH]} \\
& [K]& [$\times$10$^{13}$cm$^{-2}$]&[K]&[$\times$10$^{14}$ cm$^{-2}$]&\\
\hline
Higuchi et al. 2018&8 $-$ 21&0.5 $-$ 16& 8 $-$ 21& $<$5.2&0.04 $-$ 3.9 \\
\hline
This work, LTE&5 $-$ 17&60 $-$ 110&10 $-$ 17&0.1  $-$ 0.7& 14 $-$ 50\\
\hline
This work, LVG &8 $-$ 15 & 20 $-$ 60 & 14 $-$ 70 & 0.2 $-$ 0.4 & 3.5 $-$ 7.5\\
\hline
\end{tabular}
\end{table*} 

\subsection{What does the [CCH]/[CH$_3$OH] abundance ratio probes ? - WCCC objects, hot corinos, molecular clouds, or PDRs }

Figure \ref{f:fig_hig} reports a compilation of the CCH and CH$_3$OH total column densities observed towards WCCC objects, hot corinos, molecular clouds, and PDRs, all derived with single-dish telescopes observations. We label as a 'molecular cloud' every object where we think the emission CCH and CH$_3$OH is dominated by the non-PDR, such as pre-stellar cores, even though we know it might be not completely true. Nonetheless, they give an indication how a possible emission from the bulk or parts not illuminated by the UV would position the object on the diagram.

The figure immediately shows that the sources in OMC-2/3 have CCH and CH$_3$OH total column densities similar to those of the WCCC prototype L1527 and the PDRs. 
Since we demonstrated in Section~\ref{sec:disc-source-emi} and Section~\ref{sec:disc-map-emi} that the CCH and CH$_3$OH line emission is dominated by the extended emission rather than the source envelopes; this similarity leads us to conclude that in the specific case of OMC-2/3, the emission is in fact dominated by the PDR encircling the molecular cloud. This is fully consistent with the fact that the OMC-2/3 filament is surrounded by luminous OB stars, the Trapezium cluster and NGC 1799 HII region, namely exposed to an intense UV illumination.

On the other hand, the CCH and CH$_3$OH total column densities observed towards the sources in Perseus surveyed by \citet{higuchi2018} are definitively different from those of the PDRs, perhaps close to those in molecular clouds, and in between the PDRs and the IRAS 16293-2224 hot corino. Given the observed (rotational) temperatures measured towards the Perseus sources by \citet{higuchi2018}, similar to those measured in our OMC-2/3 sample (see Tab. \ref{tab:surveys}), it remains to be verified whether the emission is not polluted or dominated by the surrounding cloud emission. In Tab.\ref{tab:surveys}, we note that the small difference observed in the derived column densities of CCH with the two different methods can be due to the sensitivity of the input parameters of the Hyperfine fitting structure Tool used for the LTE analysis (see \ref{sec:lte_cch}). Moreover, four of the Perseus sources, namely, SVS 13A, IRAS 2A, IRAS 4A and IRAS 4B, are confirmed hot corinos by interferometric observations (\citealt{bottinelli2004}, \citealt{jorgensen2005}, \citealt{sakai2006}, \citealt{bottinelli2007}, \citealt{taquet2015}, \citealt{de_simone2017}, \citealt{lopez2017},  ). In Fig.~\ref{f:fig_hig}, these sources are scattered from the hot corino prototype and their hot corino nature is not always obvious (e.g. SVS 13A, IRAS 2A) if we refer to this plot only. 

Finally, a similar word of caution may apply to the surveys by \citet{graninger2016} and \citet{lindberg2016}, given, again, the measured low (rotational) temperatures (Tab. \ref{tab:surveysold}). Besides, the methanol column densities in their sources are lower than $\sim 10^{13}$ \pcm, namely less than for the Perseus sources (see Fig. \ref{f:fig_hig}), which would suggest, again emission from dense cold cloud gas. 

In conclusion, our results strongly suggest that the [CCH]/[CH$_3$OH] abundance ratio derived by single-dish observations is not reliable enough to trace the envelope of the protostars. 
Since several WCCC objects and hot corinos have been classified solely based on single-dish observations of small carbon chains and CH$_3$OH (see Section~\ref{sec:previous-survesy}), we suggest those classifications be verified by better spatial resolution or higher excitation lines observations that would be capable of focusing on the inner protostellar envelopes and getting rid of the extended emission from the cloud and the PDR necessarily associated with the targeted sources.


\section{Conclusions} \label{sec:conclusion}
We carried out CCH and CH$_3$OH observations with the single-dish IRAM-30m and the Nobeyama-45m telescopes of nine Solar-like protostars located in the OMC-2/3 filament in order to determine their chemical nature.
Here are our main conclusions:
\begin{enumerate}[label=\arabic*]

\item The derived gas temperatures and gas densities of CCH and CH$_3$OH towards the six sources out of the nine sources and the results from the map observations indicates that the two species are emitted from an external layer not associated with the protostars.

\item The column density ratios [CCH]/[CH$_3$OH], from 1.5 to 13.5 within the error bars, derived for our source sample specify that this external layer is likely the PDR. 

\item We could not achieve the first goal of our study that is finding hot corino or WCCC candidates in the OMC-2/3 filament. The abundance ratio used is not reliable for single-dish observations. We thus need to choose other tracers for single-dish observations or to employ interferometric observations to avoid contamination from the cloud and/or the associated PDR.
\end{enumerate}
   
\begin{acknowledgements}
We thank the referee for the careful reading and the constructive remarks that helped to improve the paper. This project has received funding from the European Research Council (ERC) under the European Union's Horizon 2020 research and innovation programme, for the Project ''The Dawn of Organic Chemistry" (DOC), grant agreement No 741002 and from the European Union's Horizon 2020 research and innovation programme under the Marie Sk\l odowska-Curie grant agreement No 811312. This work is based on observations carried out under the project numbers D01-18  and p279673 with the IRAM-30m telescope and based on observations carried out with the Nobeyama-45m . IRAM is supported by INSU/CNRS (France), MPG (Germany) and IGN (Spain). The Nobeyama 45-m radio telescope is operated by Nobeyama Radio Observatory, a branch of National Astronomical Observatory of Japan. 

\end{acknowledgements}

\bibliographystyle{aa}
\bibliography{refs}

\appendix
\section{Sample of lines}\label{appdxA}
Sample of CCH and CH$_3$OH lines detected at 1.3 mm (IRAM-30m) and at 3 mm (Nobeyama-45m) for each source (except for SIMBA-a, FIR6c and FIR2 as explained in Sect. 4.1 and 5.1.).
\newpage
\begin{figure*}[ht]
\centering
\includegraphics[width=17cm]{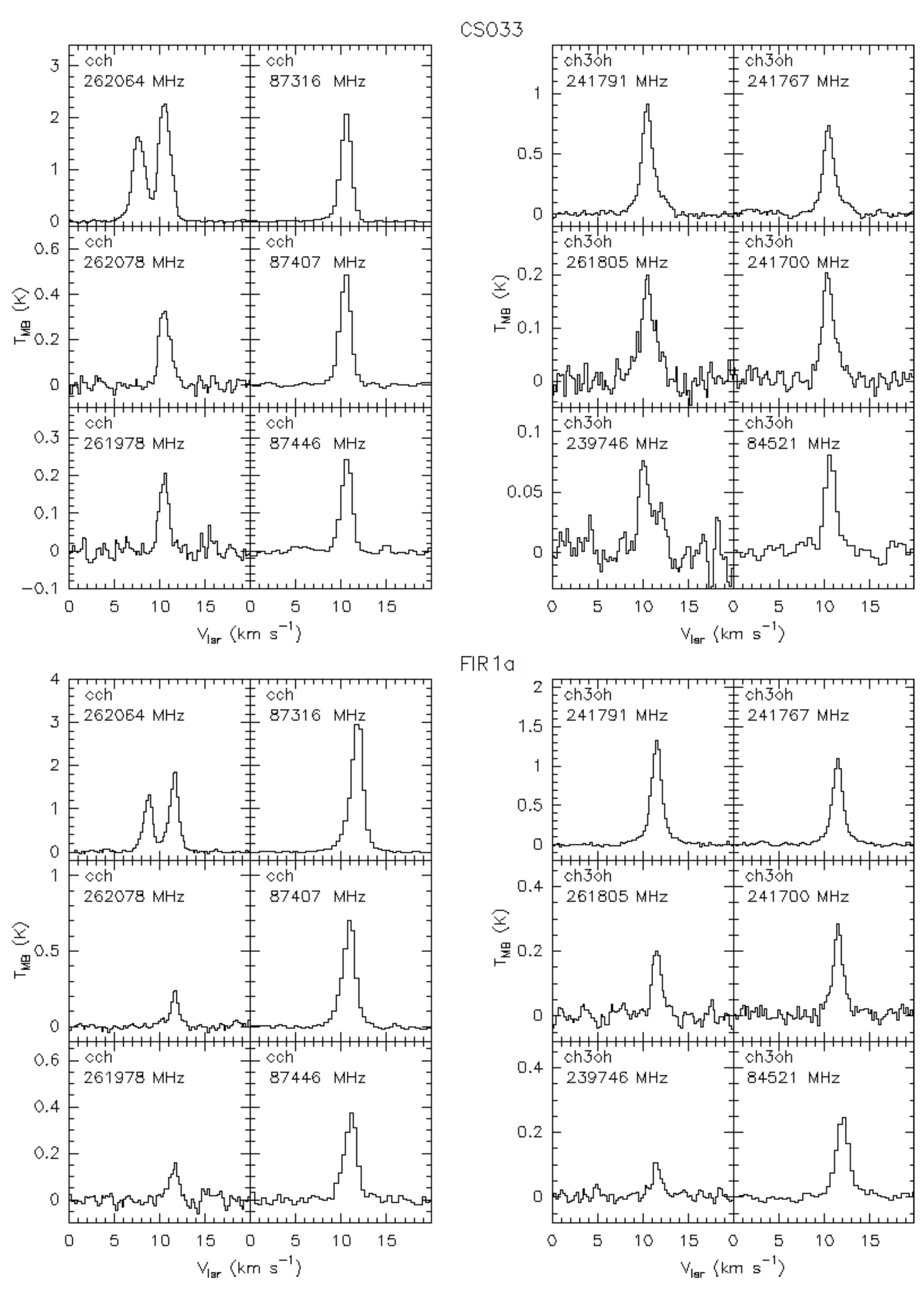}
\caption{Sample of CCH and CH$_3$OH lines of the sources CSO33 (\textit{top}) and FIR1a (\textit{Bottom})}
\end{figure*}

\begin{figure*}[!ht]
\includegraphics[width=17cm]{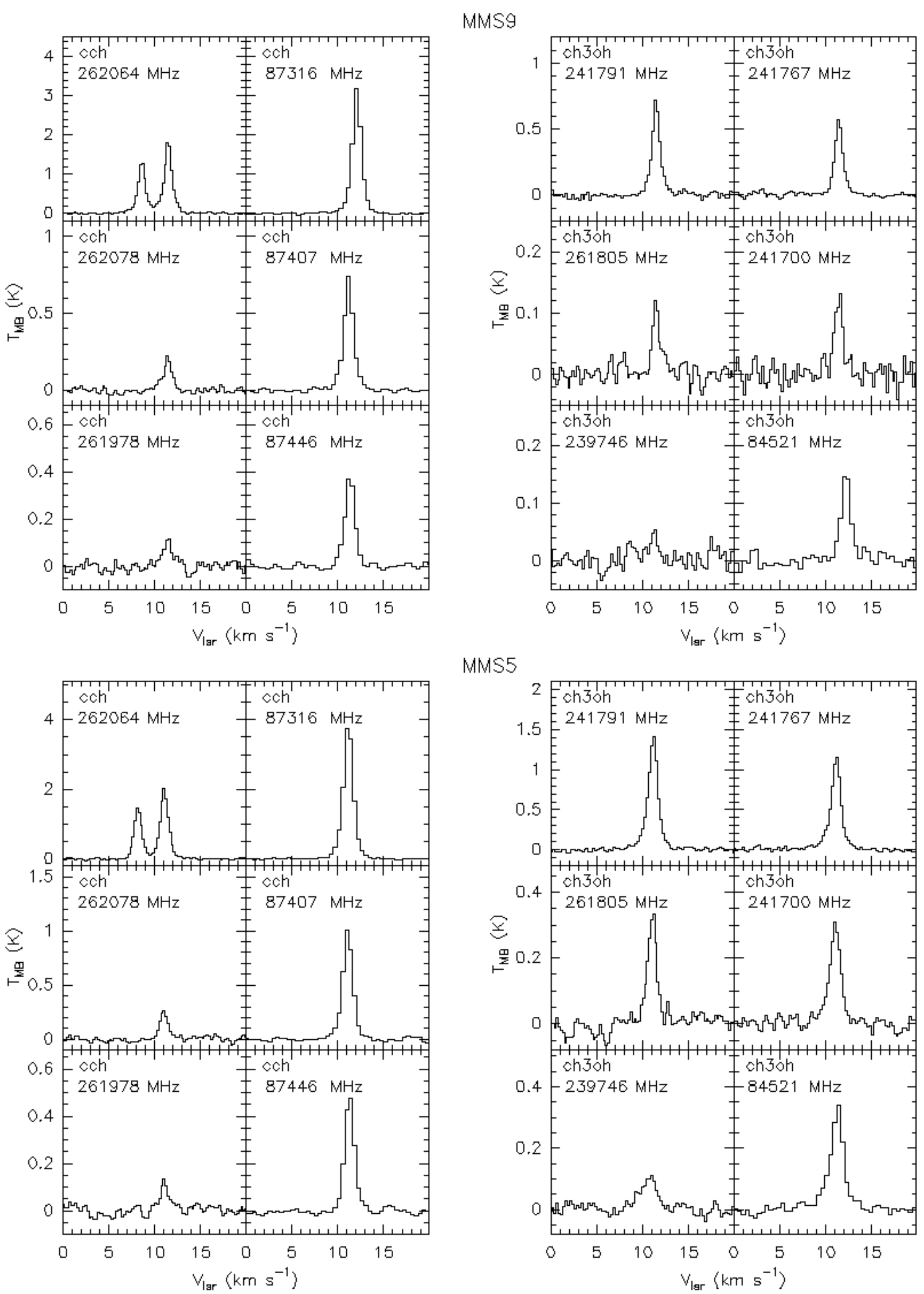}
\caption{Sample of CCH and CH$_3$OH lines of the sources MMS9 (\textit{top}) and MMS5 (\textit{Bottom})}
\end{figure*}
\begin{figure*}[!ht]
\includegraphics[width=17cm]{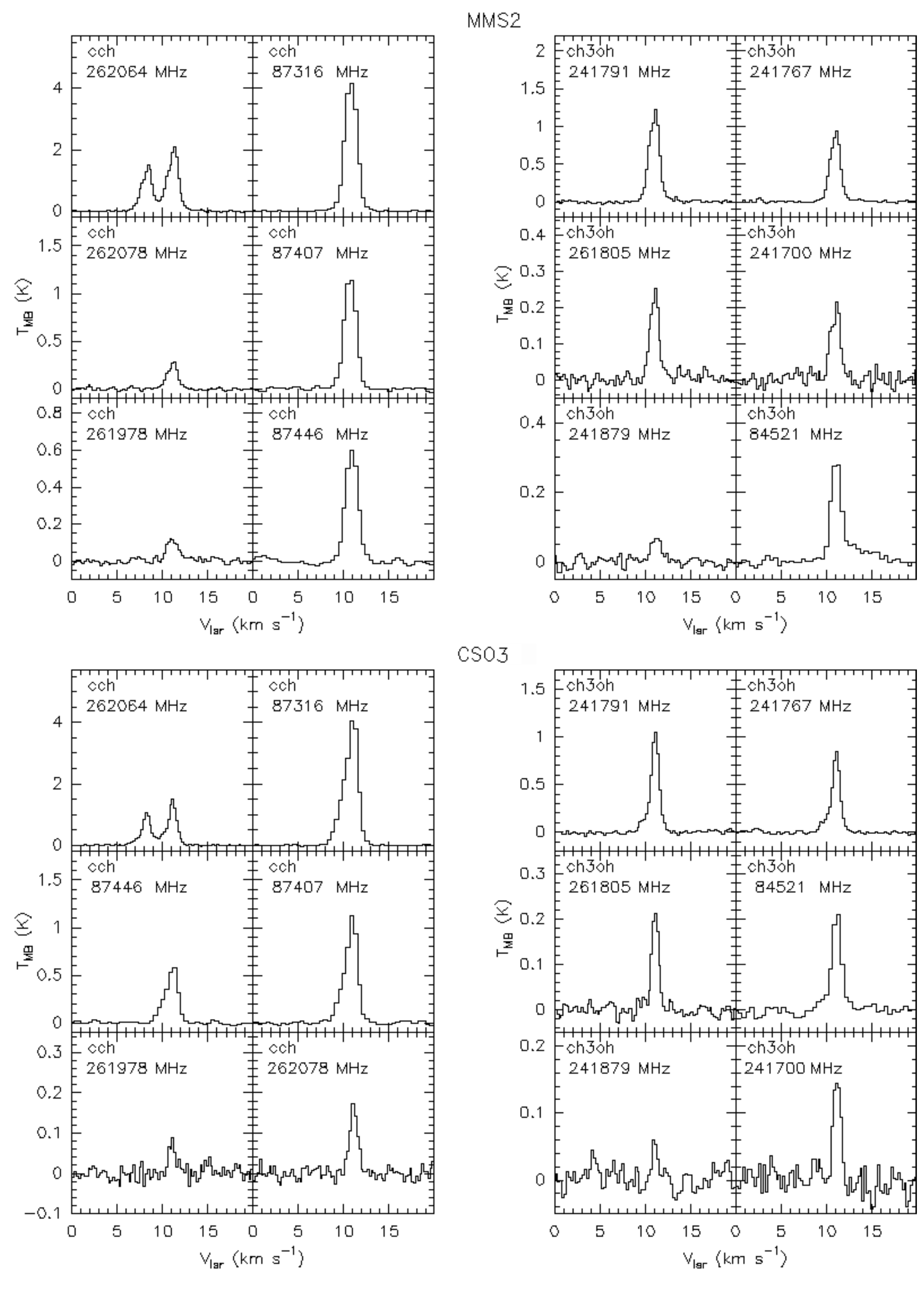}
\caption{Sample of CCH and CH$_3$OH lines of the sources MMS2 (\textit{top}) and CSO3 (\textit{Bottom})}
\end{figure*}
\begin{figure*}[!ht]
\includegraphics[width=\linewidth]{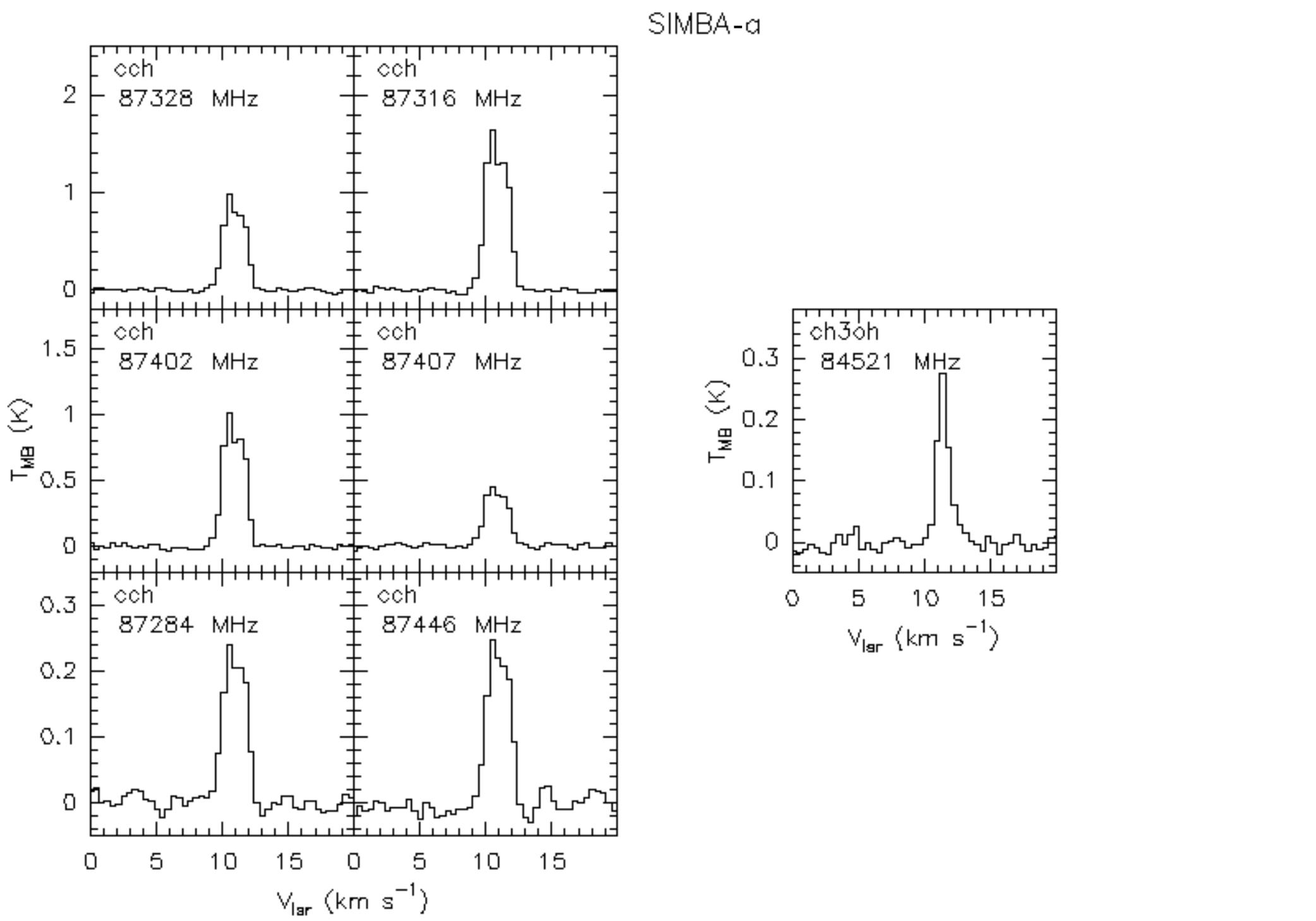}
\caption{Sample of CCH and CH$_3$OH lines of the source SIMBA-a}
\end{figure*}

\begin{figure*}[!ht]
\centering
\includegraphics[width=0.8\linewidth]{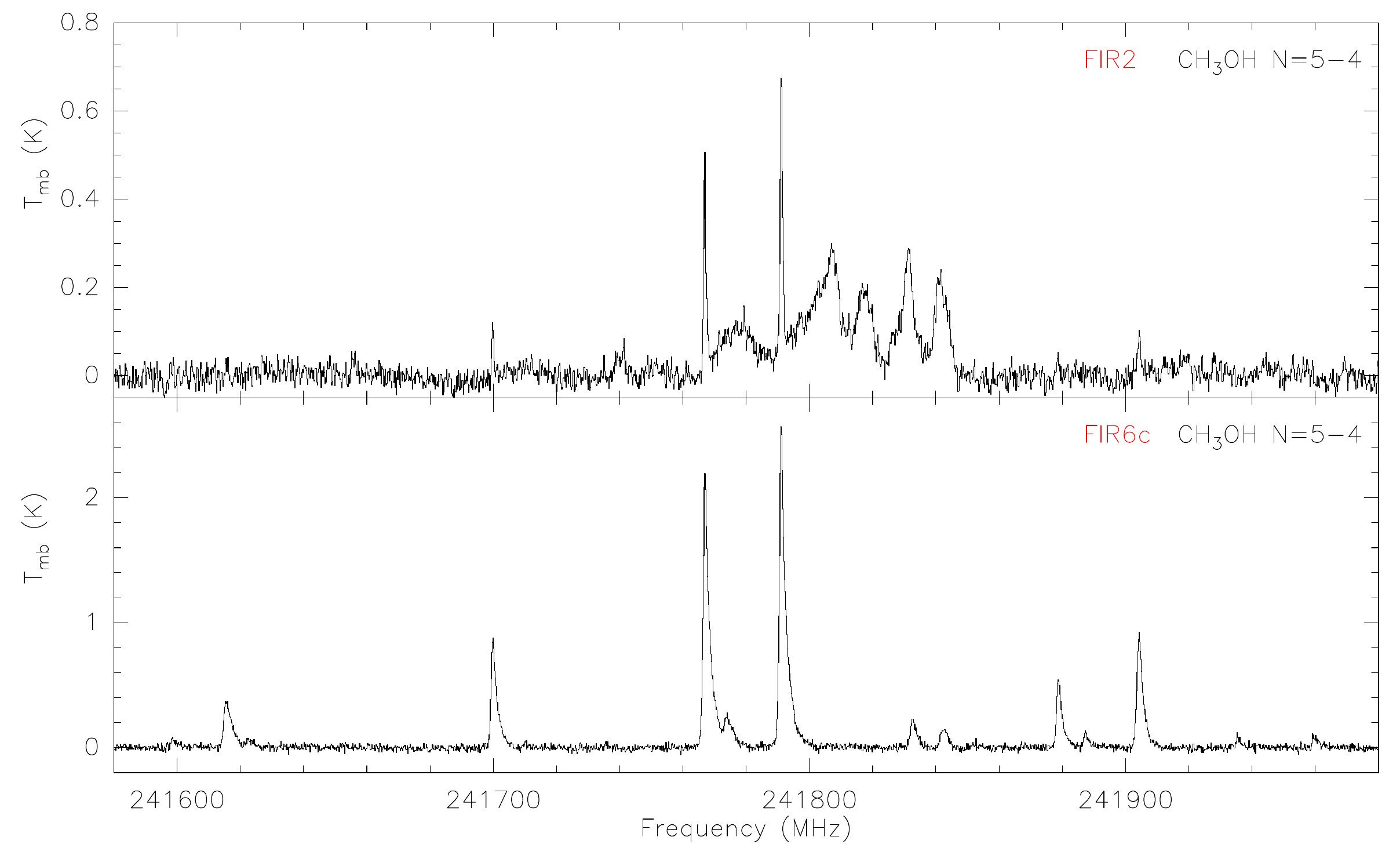}
\label{fig:c4_c5_spectra}
\caption{Lines of the $N$=5-4 transition of CH$_3$OH for FIR2 (\textit{top}) and FIR6c (\textit{Bottom}). In both spectra, broad shoulders indicating the presence of outflows are present.}
\end{figure*}

\FloatBarrier

\section{Gaussian fit results}
\subsection{Single-pointing}\label{appdxB1}
Here we present the Gaussian fits performed on each of the detected lines of CCH and CH$_3$OH for each sources. Calibration errors of 20\% and 40\% for the IRAM-30m and Nobeyama-45m observations have not been included.

\begin{table}[ht]
\caption{Gaussian fit parameters for narrow component in CSO33.}
\label{tab:fitN-c1}
\centering
\setlength{\tabcolsep}{0.07cm}
\begin{tabular}{|c|ccccc|}
\hline
\multirowthead{2}{Molecule}&\multirowthead{2}{Frequency} & Area& $V_{\text{lsr}}$  & FWHM  & T$_{\text{peak}}$  \\
 &&[mK.km/s] & [km/s] &[km/s] &[mK] \\
 \hline
\multirow{8}{*}{CH$_3$OH}&\ \ 84.521 & 122  $\pm$ 10&10.7  $\pm$ 0.2 & 1.4 $\pm$  0.2&80  $\pm$ 10  \\
&239.746 & 146 $\pm$ 18 & 10.6 $\pm$ 0.1 & 2.5 $\pm$ 0.3 & 74 $\pm$ 14 \\
&241.700 & 114 $\pm$ 40 & 10.3 $\pm$ 0.1 & 0.9 $\pm$ 0.2 & 202 $\pm$ 13 \\
&241.767 & 668 $\pm$ 44 & 10.5 $\pm$ 0.1 & 1.1 $\pm$ 0.1 & 569 $\pm$ 17 \\
&241.791 & 740 $\pm$ 42 & 10.4 $\pm$ 0.1 & 1.1 $\pm$ 0.1 & 656 $\pm$ 15 \\
&241.879 & 171 $\pm$ 16 & 10.3 $\pm$ 0.1 & 1.4 $\pm$ 0.2 & 117 $\pm$ 16 \\
&243.915 & 155 $\pm$ 14 & 10.4 $\pm$ 0.1 & 2.1 $\pm$ 0.2 & 70 $\pm$ 13 \\
&261.805 & 268 $\pm$ 20 & 10.5 $\pm$ 0.1 & 1.9 $\pm$ 0.1 & 198 $\pm$ 19 \\
\hline
\multirow{14}{*}{CCH}&\ \ 87.284 &333 $\pm$  10 &10.6 $\pm$ 0.2  &1.4 $\pm$ 0.2  &224 $\pm$ 10 \\
&\ \ 87.316 &2614 $\pm$ 30  &10.6 $\pm$  0.2 &1.3 $\pm$  0.2 & 1910 $\pm$ 10 \\
&\ \ 87.328 & 1397 $\pm$  10 &10.6 $\pm$  0.2 &1.3 $\pm$  0.2 &1030 $\pm$ 10 \\
&\ \ 87.402 &1405 $\pm$  60 &10.6 $\pm$   0.2&1.3 $\pm$  0.2 &1050 $\pm$ 10 \\
&\ \ 87.407 &599 $\pm$ 90  &10.6 $\pm$  0.2 &1.3 $\pm$ 0.2  & 430 $\pm$ 10\\
&\ \ 87.446 &367 $\pm$ 10  &10.6 $\pm$ 0.2  &1.4 $\pm$  0.2 & 244 $\pm$  10\\
&261.978 & 256 $\pm$ 14 & 10.5 $\pm$ 0.1 & 1.2 $\pm$ 0.1 & 195 $\pm$ 18 \\
&262.004 & 4868 $\pm$ 50 & 10.5 $\pm$ 0.1 & 1.5 $\pm$ 0.1 & 3020 $\pm$ 17 \\
&262.006 & 3902 $\pm$ 70 & 10.5 $\pm$ 0.1 & 1.5 $\pm$ 0.1 & 2470 $\pm$ 19 \\
&262.064 & 3677 $\pm$ 34 & 10.5 $\pm$ 0.1 & 1.5 $\pm$ 0.1 & 2260 $\pm$ 14 \\
&262.067 & 2334 $\pm$ 15 & 10.5 $\pm$ 0.1 & 1.4 $\pm$ 0.1 & 1520 $\pm$ 17 \\
&262.078 & 504 $\pm$ 18 & 10.5 $\pm$ 0.1 & 1.4 $\pm$ 0.1 & 332 $\pm$ 21 \\
&262.208 & 526 $\pm$ 16 & 10.5 $\pm$ 0.1 & 1.5 $\pm$ 0.1 & 332 $\pm$ 18 \\
&262.250 & 180 $\pm$ 14 & 10.5 $\pm$ 0.1 & 1.3 $\pm$ 0.1 & 137 $\pm$ 16\\
\hline
\end{tabular}
\end{table}

\begin{table}[ht]
\centering
\caption{Gaussian fit parameters for narrow component in FIR1a.}
\label{tab:fitN-c6}
\setlength{\tabcolsep}{0.1cm}
\begin{tabular}{|c|ccccc|}
\hline
\multirowthead{2}{Molecule}&\multirowthead{2}{Frequency} & Area& $V_{\text{lsr}}$  & FWHM  & T$_{\text{peak}}$  \\
 &&[mK.km/s]  & [km/s] & [km/s]  & [mK]  \\
 \hline
\multirow{8}{*}{CH$_3$OH}&\ \ 84.521 &411 $\pm$ 10  &11.7  $\pm$  0.2&1.7 $\pm$   0.2& 229 $\pm$ 10\\
&239.746 & 123 $\pm$ 13 & 11.6 $\pm$ 0.1 & 1.1 $\pm$ 0.2 & 102 $\pm$ 13 \\
&241.700 & 290 $\pm$ 14 & 11.6 $\pm$ 0.1 & 1.1 $\pm$ 0.1 & 230 $\pm$ 16 \\
&241.767 & 1020 $\pm$ 60 & 11.5 $\pm$ 0.1 & 1.1 $\pm$ 0.1 & 853 $\pm$ 17 \\
&241.791 & 1260 $\pm$ 50 & 11.5 $\pm$ 0.1 & 1.1 $\pm$ 0.1 & 1050 $\pm$ 14 \\
&241.879 & 162 $\pm$ 12 & 11.6 $\pm$ 0.1 & 1.3 $\pm$ 0.1 & 119 $\pm$ 14\\
&243.915 & 182 $\pm$ 18 & 11.7 $\pm$ 0.1 & 1.6 $\pm$ 0.2 & 105 $\pm$ 16 \\
&261.805 & 265 $\pm$ 16 & 11.5 $\pm$ 0.1 & 1.2 $\pm$ 0.1 & 203 $\pm$ 20 \\
\hline
\multirow{14}{*}{CCH}&\ \ 87.284 &346 $\pm$ 116  &11.4 $\pm$  0.2 &1.3 $\pm$  0.2 & 131 $\pm$ 10 \\
&\ \ 87.316 &2910 $\pm$  33 &11.3 $\pm$   0.2&1.2 $\pm$   0.2&2260 $\pm$ 11 \\
&\ \ 87.328 &1210 $\pm$  103 &11.3 $\pm$  0.2 &1.1 $\pm$  0.2 &1030 $\pm$ 11 \\
&\ \ 87.402 &1597 $\pm$ 10  &11.1 $\pm$   0.2&1.2 $\pm$ 0.2   &1270 $\pm$ 14  \\
&\ \ 87.407 &890 $\pm$  88 &11.1 $\pm$   0.2&1.0 $\pm$  0.2 &388 $\pm$ 10  \\
&\ \ 87.446 &312 $\pm$ 18  &11.3 $\pm$  0.2 &1.2 $\pm$ 0.2  &249 $\pm$  13\\
&261.978 & 140 $\pm$ 17 & 11.6 $\pm$ 0.1 & 1.1 $\pm$ 0.1 & 147 $\pm$ 21 \\
&262.004 & 2435 $\pm$ 28 & 11.5 $\pm$ 0.1 & 1.0 $\pm$ 0.1 & 2230 $\pm$ 23 \\
&262.006 & 1714 $\pm$ 35 & 11.5 $\pm$ 0.1 & 1.0 $\pm$ 0.1 & 1560 $\pm$ 20 \\
&262.064 & 1841 $\pm$ 31 & 11.6 $\pm$ 0.1 & 1.0 $\pm$ 0.1 & 1680 $\pm$ 28 \\
&262.067 & 1184 $\pm$ 36 & 11.5 $\pm$ 0.1 & 1.0 $\pm$ 0.1 & 1060 $\pm$ 25 \\
&262.078 & 210 $\pm$ 42 & 11.6 $\pm$ 0.1 & 1.0 $\pm$ 0.2 & 235 $\pm$ 54 \\
&262.208 & 181 $\pm$ 18 & 11.6 $\pm$ 0.1 & 1.2 $\pm$ 0.1 & 199 $\pm$ 21 \\
&262.250 & 90 $\pm$ 10 & 11.8 $\pm$ 0.2 & 0.8 $\pm$ 0.1 & 103 $\pm$ 21\\
\hline
\end{tabular}
\end{table}

\begin{table}[ht]
\centering
\caption{Gaussian fit parameters for narrow component in MMS9.}
\label{tab:fitN-c7}
\setlength{\tabcolsep}{0.1cm}
\begin{tabular}{|c|ccccc|}
\hline
\multirowthead{2}{Molecule}&\multirowthead{2}{Frequency} & Area& $V_{\text{lsr}}$  & FWHM  & T$_{\text{peak}}$  \\
 &&[mK.km/s]  & [km/s] & [km/s]  & [mK]  \\
 \hline
\multirow{8}{*}{CH$_3$OH}&84.521 & 213$\pm$ 10  &11.5  $\pm$ 0.2 &1.3 $\pm$ 0.2  &154 $\pm$  10\\
&239.746 & 47 $\pm$ 9 & 11.2 $\pm$ 0.1 & 0.8 $\pm$ 0.2 & 52 $\pm$ 13 \\
&241.700 &150 $\pm$ 15 & 11.4 $\pm$ 0.1 & 1.1 $\pm$ 0.1 & 132 $\pm$ 19 \\
&241.767 & 220 $\pm$ 12 & 11.4 $\pm$ 0.1 & 0.7 $\pm$ 0.1 & 308 $\pm$ 15 \\
&241.791 & 300 $\pm$ 60 & 11.4 $\pm$ 0.1 & 0.7 $\pm$ 0.1 & 430 $\pm$ 19 \\
&241.879 & 57 $\pm$ 9 & 11.4 $\pm$ 0.1 & 0.7 $\pm$ 0.1 & 74 $\pm$ 13 \\
&243.915 & 105 $\pm$ 12 & 11.5 $\pm$ 0.1 & 1.8 $\pm$ 0.2 & 55 $\pm$ 12 \\
&261.805 & 115 $\pm$ 12 & 11.4 $\pm$ 0.1 & 0.8 $\pm$ 0.1 & 130 $\pm$ 16 \\
\hline
\multirow{14}{*}{CCH}&87.284 &471 $\pm$  15 &11.6 $\pm$ 0.2  &1.3 $\pm$  0.2 &343 $\pm$ 13 \\
&87.316 &2046 $\pm$ 10&11.6 $\pm$ 0.2  &0.9 $\pm$ 0.2   &2230 $\pm$ 12 \\
&87.328 &1128 $\pm$  118 &11.6 $\pm$ 0.2  &0.9 $\pm$ 0.2  &1200 $\pm$  12\\
&87.402 &1494 $\pm$  32 &11.6 $\pm$  0.2 & 0.9 $\pm$ 0.2  &1510 $\pm$ 13 \\
&87.407 &463 $\pm$ 79  &11.6 $\pm$  0.2 &0.9 $\pm$  0.2 &483 $\pm$ 12 \\
&87.446 &519 $\pm$ 11 &11.5 $\pm$  0.2 &1.3 $\pm$ 0.2  &378 $\pm$  11\\
&261.978 & 70 $\pm$ 30 & 11.5 $\pm$ 0.1 & 0.6 $\pm$ 0.2 & 123 $\pm$ 16 \\
&262.004 & 1830 $\pm$ 24 & 11.4 $\pm$ 0.1 & 0.8 $\pm$ 0.1 & 2030 $\pm$ 21 \\
&262.006 & 1370 $\pm$ 16 & 11.4 $\pm$ 0.1 & 0.9 $\pm$ 0.1 & 1520 $\pm$ 17 \\
&262.064 & 1500 $\pm$ 12 & 11.5 $\pm$ 0.1 & 0.9 $\pm$ 0.1 & 1610 $\pm$ 19 \\
&262.067 & 990 $\pm$ 10 & 11.4 $\pm$ 0.1 & 0.9 $\pm$ 0.1 & 1100 $\pm$ 16 \\
&262.078 & 214 $\pm$ 13 & 11.4 $\pm$ 0.1 & 1.0 $\pm$ 0.1 & 210 $\pm$ 18 \\
&262.208 & 197 $\pm$ 12 & 11.4 $\pm$ 0.1 & 0.8 $\pm$ 0.1 & 235 $\pm$ 17 \\
&262.250 & 68 $\pm$ 11 & 11.5 $\pm$ 0.1 & 0.7 $\pm$ 0.1 & 91 $\pm$ 18\\
\hline
\end{tabular}
\end{table}

\begin{table}[ht]
\centering
\caption{Gaussian fit parameters for narrow component in MMS5.}
\label{tab:fitN-c3}
\setlength{\tabcolsep}{0.07cm}
\begin{tabular}{|c|ccccc|}
\hline
\multirowthead{2}{Molecule}&\multirowthead{2}{Frequency} & Area& $V_{\text{lsr}}$  & FWHM  & T$_{\text{peak}}$  \\
 && [mK.km/s]  & [km/s] & [km/s]  &  [mK]  \\
 \hline
\multirow{9}{*}{CH$_3$OH}&\ \ 84.521 &245 $\pm$ 60  &11.4  $\pm$ 0.2 &1.1 $\pm$ 0.2  &216 $\pm$  10\\
&239.746 & 207 $\pm$ 19 & 10.8 $\pm$ 0.1 & 1.8 $\pm$ 0.2 & 108 $\pm$ 17 \\
&241.700 & 250 $\pm$ 90 & 11.1 $\pm$ 0.1 & 1.1 $\pm$ 0.1 & 216 $\pm$ 17 \\
&241.767 & 880 $\pm$ 70 & 11.2 $\pm$ 0.1 & 0.9 $\pm$ 0.1 & 880 $\pm$ 21 \\
&241.791 & 1020 $\pm$ 40 & 11.2 $\pm$ 0.1 & 0.9 $\pm$ 0.1 & 1020 $\pm$ 19 \\
&241.879 & 280 $\pm$ 20 & 11.0 $\pm$ 0.1 & 1.6 $\pm$ 0.1 & 163 $\pm$ 20 \\
&241.887 & 60 $\pm$ 10 & 11.2 $\pm$ 0.1 & 0.9 $\pm$ 0.2 & 60 $\pm$ 16 \\
&243.915 &264 $\pm$ 17 & 11.0 $\pm$ 0.1 & 1.6 $\pm$ 0.1 & 151 $\pm$ 18 \\
&261.805 & 416 $\pm$ 33 & 11.1 $\pm$ 0.1 & 1.2 $\pm$ 0.1 & 320 $\pm$ 27 \\
\hline
\multirow{15}{*}{CCH}&\ \ 87.284 &670 $\pm$ 11   &11.2 $\pm$  0.2 &1.1 $\pm$  0.2 &477 $\pm$  10\\
&\ \ 87.316 &4184 $\pm$ 12  &11.2 $\pm$ 0.2   &1.0 $\pm$ 0.2  &3450 $\pm$ 13 \\
&\ \ 87.328 &2334 $\pm$ 20  &11.2 $\pm$ 0.2  &1.0 $\pm$  0.2 &1930 $\pm$ 11  \\
&\ \ 87.402 &2132 $\pm$ 26 &11.2 $\pm$  0.2 &0.9 $\pm$ 0.2  &1910 $\pm$ 17 \\
&\ \ 87.407 &1041 $\pm$ 90  &11.1 $\pm$ 0.2  &1.1 $\pm$  0.2 &862 $\pm$  15\\
&\ \ 87.446 &654 $\pm$ 13  &11.3 $\pm$  0.2 &1.1 $\pm$ 0.2  &488 $\pm$ 12 \\
&261.978 & 130 $\pm$ 16 & 11.0 $\pm$ 0.1 & 1.0 $\pm$ 0.2 & 124 $\pm$ 19 \\
&262.004 & 2582 $\pm$ 36 & 11.0 $\pm$ 0.1 & 1.0 $\pm$ 0.1 & 2360 $\pm$ 21 \\
&262.006 & 2104 $\pm$ 22 & 11.0 $\pm$ 0.1 & 1.0 $\pm$ 0.1 & 1970 $\pm$ 21 \\
&262.064 & 1601 $\pm$ 10 & 11.0 $\pm$ 0.1 & 0.9 $\pm$ 0.1 & 1610 $\pm$ 23 \\
&262.067 & 1556 $\pm$ 70 & 11.0 $\pm$ 0.1 & 1.0 $\pm$ 0.1 & 1410 $\pm$ 23 \\
&262.078 & 289 $\pm$ 15 & 11.0 $\pm$ 0.1 & 1.1 $\pm$ 0.1 & 258 $\pm$ 20 \\
&262.208 & 274 $\pm$ 13 & 11.0 $\pm$ 0.2 & 1.0 $\pm$ 0.1 & 252 $\pm$ 18 \\
&262.236 & 79 $\pm$ 15 & 10.7 $\pm$ 0.1 & 1.0 $\pm$ 0.2 & 73 $\pm$ 16\\
&262.250 & 86 $\pm$ 17 & 11.0 $\pm$ 0.1 & 0.7 $\pm$ 0.2 & 118 $\pm$ 24 \\
\hline
\end{tabular}
\end{table}

\begin{table}[ht]
\centering
\caption{Gaussian fit parameters for narrow component in MMS2.}
\label{tab:fitN-c8}
\setlength{\tabcolsep}{0.07cm}
\begin{tabular}{|c|ccccc|}
\hline
\multirowthead{2}{Molecule}&\multirowthead{2}{Frequency} & Area& $V_{\text{lsr}}$  & FWHM  & T$_{\text{peak}}$  \\
 &&[mK.km/s]  & [km/s] & [km/s]  & [mK]  \\
 \hline
\multirow{8}{*}{CH$_3$OH}&\ \ 84.521 &362 $\pm$  15 & 11.1 $\pm$ 0.2 &1.2 $\pm$ 0.2  &273 $\pm$ 9 \\
&239.746 & 86 $\pm$ 13 & 11.0 $\pm$ 0.1 & 1.7 $\pm$ 0.3 & 48 $\pm$ 12 \\
&241.700 & 269 $\pm$ 13 & 11.0 $\pm$ 0.1 & 1.3 $\pm$ 0.1 & 199 $\pm$ 16 \\
&241.767 & 1220 $\pm$ 20 & 11.0 $\pm$ 0.1 & 1.3 $\pm$ 0.1 & 901 $\pm$ 13 \\
&241.791 & 1603 $\pm$ 10 & 11.0 $\pm$ 0.1 & 1.3 $\pm$ 0.1& 1160 $\pm$ 13 \\
&241.879 & 97 $\pm$ 12 & 11.2 $\pm$ 0.1 & 1.3 $\pm$ 0.2 &72 $\pm$ 14 \\
&243.915 & 139 $\pm$ 11 & 11.2 $\pm$ 0.1 & 1.5 $\pm$ 0.2 & 71 $\pm$ 12 \\
&261.805 & 298 $\pm$ 12 & 11.0 $\pm$ 0.1 & 1.2 $\pm$ 0.1 & 239 $\pm$ 15 \\
\hline
\multirow{13}{*}{CCH}&\ \ 87.284 &957 $\pm$ 16  &10.9 $\pm$  0.2 &1.6 $\pm$ 0.2  &578 $\pm$ 11 \\
&\ \ 87.316 &7084 $\pm$  15 &10.8 $\pm$   0.2&1.6 $\pm$  0.2 &4290 $\pm$ 12 \\
&\ \ 87.328 &3981 $\pm$ 17  &10.9 $\pm$ 0.2  &1.5 $\pm$ 0.2  &2480 $\pm$ 14  \\
&\ \ 87.402 &4165 $\pm$  18 &10.8 $\pm$ 0.2  &1.5 $\pm$  0.2 &2610 $\pm$ 15 \\
&\ \ 87.407 &1930 $\pm$  15 &10.7 $\pm$  0.2 &1.5 $\pm$ 0.2  &1190 $\pm$ 15 \\
&\ \ 87.446 &1021 $\pm$  16 &10.9 $\pm$  0.2 &1.6 $\pm$  0.2 & 608 $\pm$ 11\\
&261.978 & 188 $\pm$ 13 & 11.1 $\pm$ 0.1 & 1.5 $\pm$ 0.1 & 120 $\pm$ 16 \\
&262.004 & 3927 $\pm$ 21 & 11.1 $\pm$ 0.1 & 1.4 $\pm$ 0.1 & 2570 $\pm$ 23 \\
&262.006 & 3334 $\pm$ 16 & 11.2 $\pm$ 0.1 & 1.6 $\pm$ 0.1 & 2000 $\pm$ 16 \\
&262.064 & 2996 $\pm$ 16 & 11.2 $\pm$ 0.1 & 1.4 $\pm$ 0.1 & 1940 $\pm$ 18 \\
&262.067 & 2203 $\pm$ 15 & 11.2 $\pm$ 0.1 & 1.5 $\pm$ 0.1 & 1360 $\pm$ 16 \\
&262.078 & 360 $\pm$ 20 & 11.1 $\pm$ 0.1 & 1.3 $\pm$ 0.1 & 257 $\pm$ 14 \\
&262.208 & 353 $\pm$ 12 & 11.1 $\pm$ 0.1 & 1.4 $\pm$ 0.1 &241 $\pm$ 14 \\
&262.250 & 129 $\pm$ 13 & 11.2 $\pm$ 0.1 & 1.3 $\pm$ 0.1 & 91 $\pm$ 17\\
\hline
\end{tabular}
\end{table}

\begin{table}[ht]
\centering
\caption{Gaussian fit parameters for narrow component in CSO3.}
\label{tab:fitN-c9}
\setlength{\tabcolsep}{0.1cm}
\begin{tabular}{|c|ccccc|}
\hline
\multirowthead{2}{Molecule}&\multirowthead{2}{Frequency} & Area& $V_{\text{lsr}}$  & FWHM  & T$_{\text{peak}}$  \\
 &&[mK.km/s]  & [km/s] &  [km/s]  &  [mK]  \\
 \hline
\multirow{6}{*}{CH$_3$OH}&\ \ 84.521 &192 $\pm$  32 &11.2  $\pm$ 0.2 &1.0 $\pm$  0.2 &179 $\pm$ 8  \\
&241.700 & 160 $\pm$ 13 & 11.2 $\pm$ 0.1 & 1.0 $\pm$ 0.1 & 157 $\pm$ 17  \\
&241.767 & 700 $\pm$ 10 & 11.1 $\pm$ 0.1 & 0.9 $\pm$ 0.1 & 725 $\pm$ 11  \\
&241.791 & 820 $\pm$ 40 & 11.1 $\pm$ 0.1 & 0.9 $\pm$ 0.1 & 898 $\pm$ 16  \\
&241.879 & 42 $\pm$ 9 & 11.1 $\pm$ 0.1 & 0.6 $\pm$ 0.2 & 62 $\pm$ 15  \\
&261.805 & 205 $\pm$ 9 & 11.1 $\pm$ 0.1 & 0.9 $\pm$ 0.1 & 220 $\pm$ 13  \\
\hline
\multirow{13}{*}{CCH}&\ \ 87.284 &470 $\pm$ 15  &11.2 $\pm$ 0.2  &1.0 $\pm$ 0.2  &320 $\pm$ 12 \\
&\ \ 87.316 &2578 $\pm$ 50  &11.2 $\pm$ 0.2  &0.9 $\pm$  0.2 &2560 $\pm$ 15 \\
&\ \ 87.328 &1745 $\pm$ 14  &11.2 $\pm$ 0.2  &1.0 $\pm$ 0.2  &1600 $\pm$ 11 \\
&\ \ 87.402 &1647 $\pm$ 29 &11.1 $\pm$ 0.2  &1.0 $\pm$ 0.2  & 1620 $\pm$ 18\\
&\ \ 87.407 &991 $\pm$ 18  &11.1 $\pm$ 0.2  &1.1 $\pm$  0.2 &851 $\pm$ 13 \\
&\ \ 87.446 &438 $\pm$  153 &11.3 $\pm$  0.2 &1.0 $\pm$ 0.2  &401 $\pm$ 12\\
&261.978 & 72 $\pm$ 11 & 11.0 $\pm$ 0.1 & 0.8 $\pm$ 0.1 & 83 $\pm$ 14 \\
&262.004 & 1900 $\pm$ 40 & 11.1 $\pm$ 0.1 & 1.0 $\pm$ 0.1 & 1890 $\pm$ 16  \\
&262.006 & 1840 $\pm$ 11 & 11.1 $\pm$ 0.1 & 0.8 $\pm$ 0.1 & 1040 $\pm$ 12  \\
&262.064 & 1510 $\pm$ 10 & 11.1 $\pm$ 0.1 & 1.0 $\pm$ 0.1 & 1390 $\pm$ 14  \\
&262.067 & 680 $\pm$ 10 & 11.1 $\pm$ 0.1 & 0.8 $\pm$ 0.1 & 775 $\pm$ 10  \\
&262.078 & 193 $\pm$ 10 & 11.1 $\pm$ 0.1 & 1.0 $\pm$ 0.1 & 177 $\pm$ 13  \\
&262.208 & 187 $\pm$ 12 & 11.1 $\pm$ 0.1 & 1.0 $\pm$ 0.1 & 177 $\pm$ 17  \\
&262.250 & 72 $\pm$ 11 & 11.1 $\pm$ 0.1 & 1.2 $\pm$ 0.2 & 56 $\pm$ 13 \\
\hline
\end{tabular}
\end{table}

\begin{table}[ht]
\centering
\caption{Gaussian fit parameters for narrow component in SIMBA-a.}
\label{tab:fitN-c2}
\setlength{\tabcolsep}{0.1cm}
\begin{tabular}{|c|ccccc|}
\hline
\multirowthead{2}{Molecule}&\multirowthead{2}{Frequency} & Area& $V_{\text{lsr}}$  & FWHM  & T$_{\text{peak}}$  \\
 && [mK.km/s]  &[km/s] & [km/s]  &  [mK]  \\
 \hline
CH$_3$OH&84.521 &226 $\pm$ 71  &10.5  $\pm$  0.2&0.9$\pm$ 0.2  &241 $\pm$ 8  \\
\hline
\multirow{6}{*}{CCH}&87.284 &327 $\pm$ 40  &10.5 $\pm$ 0.2  &1.3 $\pm$  0.2 &233 $\pm$  13\\
&87.316 &2000 $\pm$ 50  &10.4 $\pm$  0.2 &1.1 $\pm$ 0.2   &1630 $\pm$  16\\
&87.328 &1165 $\pm$ 49  &10.5 $\pm$  0.2 &1.2 $\pm$  0.2 &950 $\pm$  15\\
&87.402 &1147 $\pm$ 46  &10.4 $\pm$  0.2 &1.1 $\pm$  0.2 &1000 $\pm$  16\\
&87.407 &554 $\pm$  64 &10.4 $\pm$  0.2 &1.2 $\pm$ 0.2  &442 $\pm$ 14  \\
&87.446 &324 $\pm$ 45 &10.6 $\pm$  0.2 &1.2 $\pm$  0.2 &247 $\pm$ 11 \\
\hline
\end{tabular}
\end{table}

\subsection{OTF Map}\label{appdxB2}
Maps of moments 1 and 2 of the lines CCH ($N$=1-0, $J$=3/2-1/2, $F$=1-0) and CH$_3$OH(2$_{-1}$ - 1$_{-1}$ E) with a threshold of 3 $\sigma$. Here we also show the intensity ratio map of two optically thin lines of methanol, (2$_{-1}$-1$_{-1}$) E and (2$_{0}$-1$_{0}$)E.

\FloatBarrier
\begin{figure*}
\centering
\includegraphics[width=0.8\linewidth]{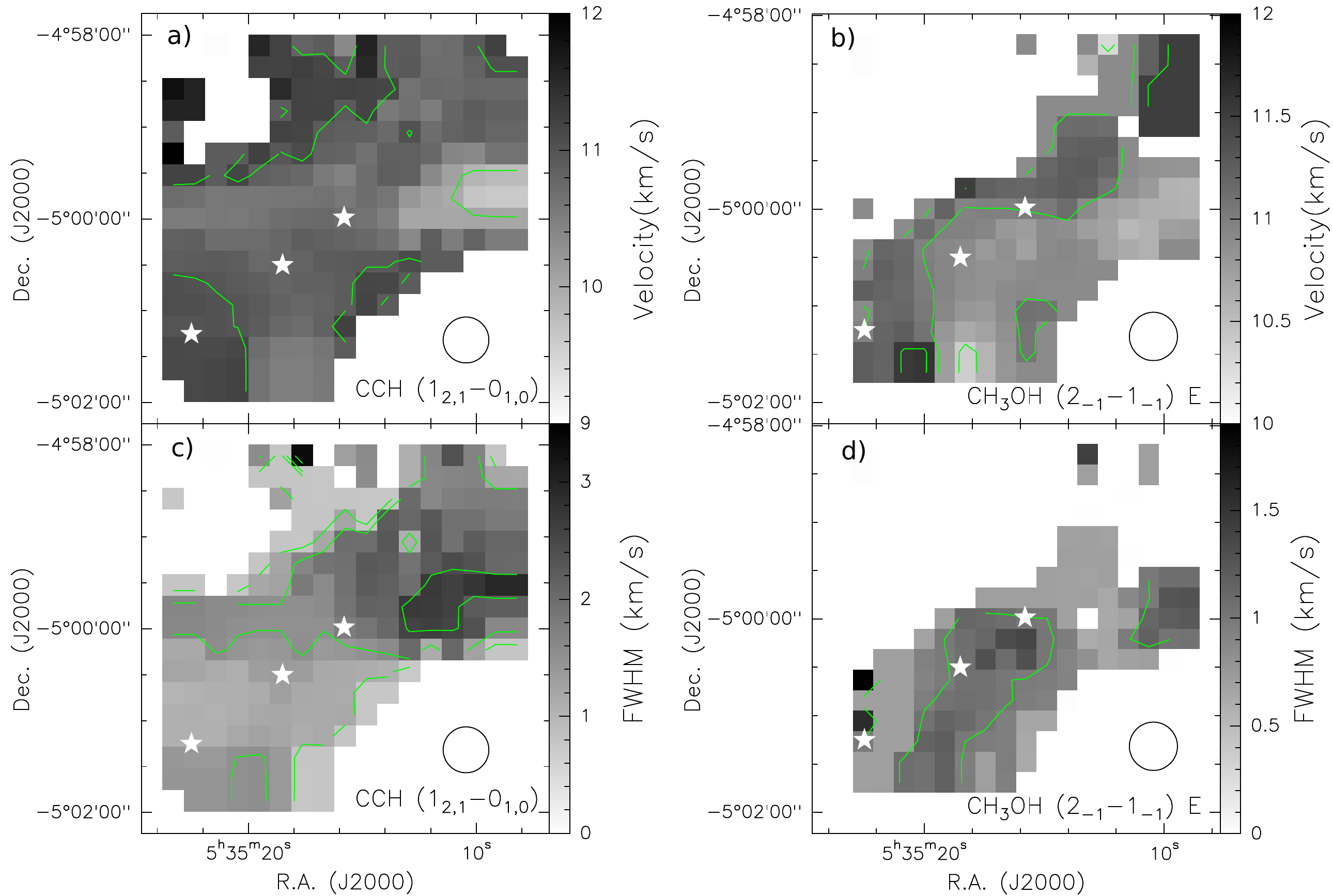}
\caption{ Moment maps of CCH and CH$_3$OH; a) and c) moments 1 and 2 of the line CCH ($N$=1-0, $J$=3/2-1/2, $F$=1-0) (a),c), respectively; b) and d) moments 1 and 2 of the line CH$_3$OH(2$_{-1}$ - 1$_{-1}$ E) respectively. The velocity integration is between 7.9 and 12.8 km/s and the threshold has been set at 3$\sigma$ for both lines. Contours levels for Moment 1 are 10 and 11 km/s for CCH and 10.5, 11 and 11.5 km/s for CH$_3$OH. Contour levels for Moments 2 are 1, 1.5 and 2.5 km/s for CCH and 1 km/s for CH$_3$OH. }
\label{fig:moments}
\end{figure*}

\begin{figure*}
\centering
\includegraphics[width=0.6\linewidth]{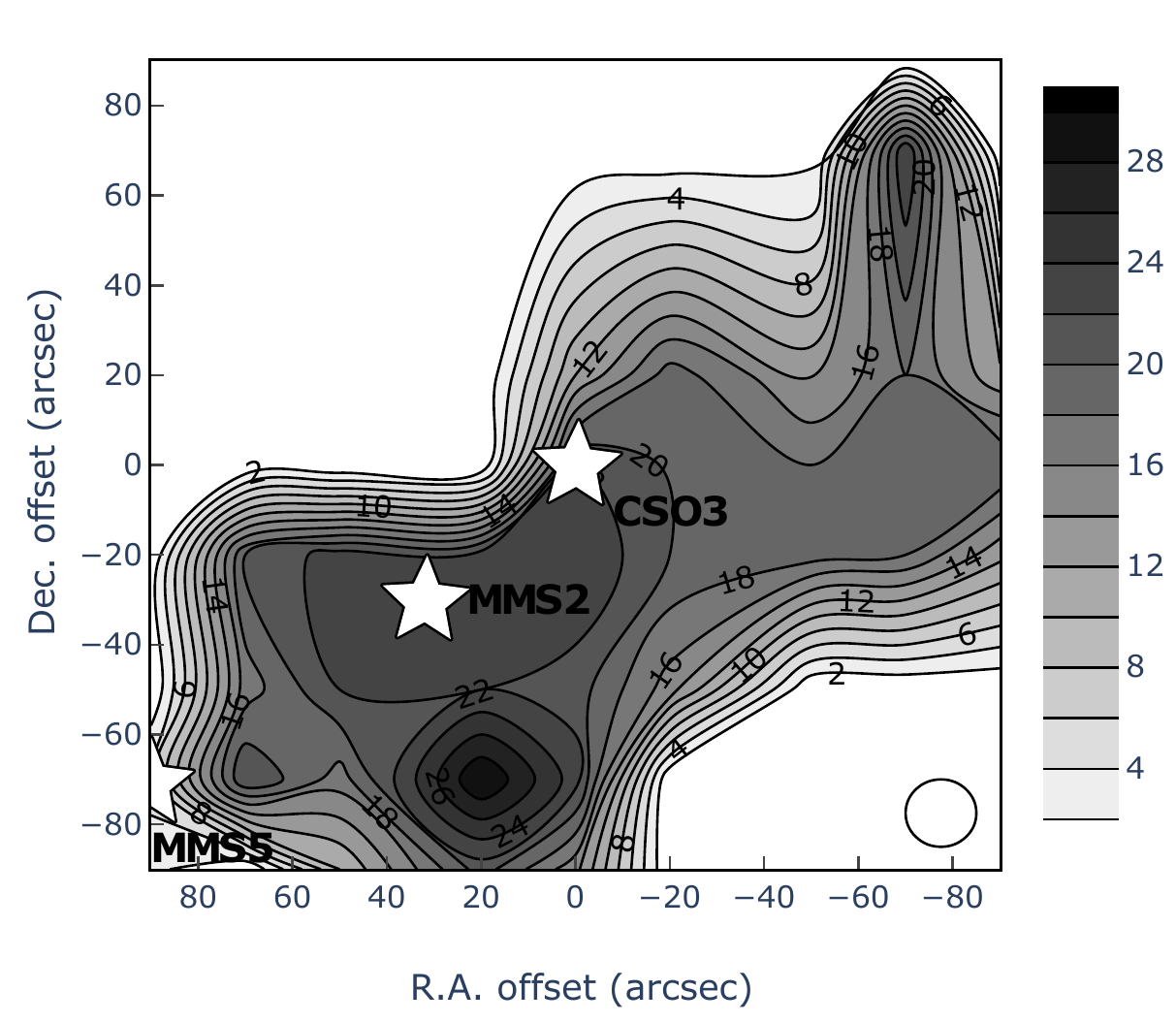}
\caption{Intensity ratio of the CH$_3$OH lines 2$_{-1}$-1$_{-1}$ E and 2$_{0}$-1$_{0}$E. The lines chosen are optically thin. We see clearly that the ratio is constant throughout the filament. Error on the line intensity ratio is less than 3.}
\label{fig:lineint_ratio}
\end{figure*}
\FloatBarrier

\section{LTE methods \& results}\label{appdxC}
Here we present the methods used to perform the LTE analysis and the results obtained.

\subsection{Single-pointing}
\subsubsection{CH$_3$OH lines}
\begin{figure*}
\includegraphics[scale=0.8]{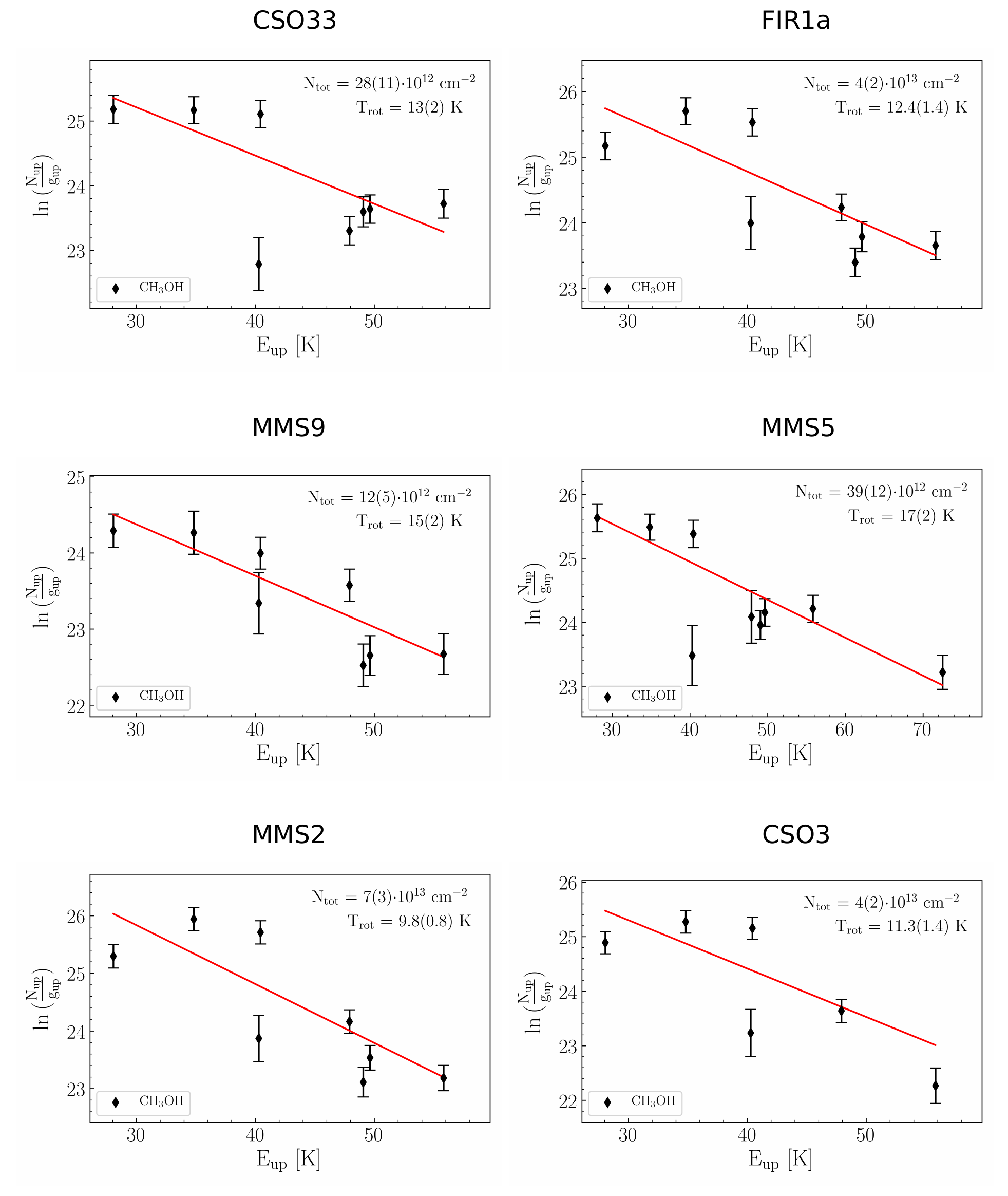}
\caption{Rotational diagrams of CH$_3$OH for each source.}
\label{rot_dg}
\end{figure*}

In the case of the methanol lines, we used the usual rotational diagram approach, which assumes optically thin lines and LTE level populations. To this end, we included all lines with a detection threshold of $ 3 \sigma$ and assumed extended emission. We  note that we verified a posteriori that the optically thin lines approximation is valid: we used the ULSA (Unbiased Line Spectral Analysis) package developed at IPAG, which is a LTE radiative transfer code to verify a posteriori this assumption. The opacity values of the CH$_3$OH lines given were less than 0.1.
The rotational diagrams are presented in Fig.~\ref{rot_dg} and the mean derived parameters are listed in Tab. \ref{tab:results_LTE}. 
In the six sources, the mean derived rotational temperatures and methanol beam-averaged column densities are 13.0 $\pm$ 1.5 K and  (4 $\pm$ 2) $\times$ 10$^{13}$ cm$^{-2}$, respectively. However, we observe a scatter of the points from the linear fit in every rotational diagram, probably due to non-LTE effects.

\begin{table*}
\centering
\makegapedcells
\setcellgapes{2pt}
\caption{Rotational temperatures and beam-averaged column densities of CH$_3$OH and CCH towards seven out of the nine target sources (Tab. \ref{tab:sources}).}
\label{tab:results_LTE}
\resizebox{\textwidth}{!}{
\begin{tabular}{|c|cc|cc|cc|cc|c|}
\hline
\multirowthead{3}{Source}& \multicolumn{2}{c|}{\text{IRAM 1mm + Nobeyama 3mm}} & \multicolumn{2}{c}{\text{IRAM 1mm}} & \multicolumn{2}{|c|}{\text{Nobeyama 3mm}} & \multicolumn{2}{c|}{\text{Mean CCH}}&Ratio\\
\cline{2-10}
 & $T_{\text{CH}_3\text{OH}}$ & $N_{\text{CH}_3\text{OH}}$& $T_{\text{CCH}}$ & $N_{\text{CCH}}$& $T_{\text{CCH}}$&$N_{\text{CCH}}$ & $T_{\text{CCH}}$&$N_{\text{CCH}}$&\multirowthead{2}{[CCH]/[CH$_3$OH]}\\
&[K]&[$\times 10^{13}$ cm$^{-2}$]&[K]&[$\times 10^{14}$ cm$^{-2}$]&[K]&[$\times 10^{14}$ cm$^{-2}$]&[K]&[$\times 10^{14}$ cm$^{-2}$]&\\
\hline
CSO33&13 $\pm$ 2&3 $\pm$ 1&11$\pm$ 1 &11.6 $\pm$ 3.5&11 $\pm$ 3&5 $\pm$ 2 & 11.0 $\pm$ 2.5 & 8.0 $\pm$ 1.5&22  $\pm$ 10   \\
FIR1a&12 $\pm$ 1&4 $\pm$ 2&17 $\pm$ 3&4 $\pm$ 1&17 $\pm$ 6&11 $\pm$ 9& 17.0 $\pm$ 4.5& 7.5 $\pm$ 5.0 & 19  $\pm$ 16 \\
MMS9&15 $\pm$ 2&1.2 $\pm$ 0.5&14 $\pm$ 2& 4 $\pm$ 1&15 $\pm$ 5&7 $\pm$ 5 & 14.5 $\pm$ 3.5&6 $\pm$ 3 & 50  $\pm$ 33 \\
MMS5&17 $\pm$ 2&4 $\pm$ 1&14 $\pm$ 2&4.5 $\pm$ 1.4&13 $\pm$ 4&10 $\pm$ 6 & 13.5 $\pm$ 3.0 & 7.0 $\pm$ 3.5& 17 $\pm$ 10 \\
MMS2& 10 $\pm$ 1& 7 $\pm$ 3&11 $\pm$ 1&8.2 $\pm$ 2.5&10 $\pm$ 3&13.7 $\pm$ 6.5&10.5 $\pm$ 2.0 & 11 $\pm$  5& 14  $\pm$ 8 \\
CSO3&11 $\pm$ 1& 4 $\pm$ 2&23 $\pm$ 4&3 $\pm$ 1&11 $\pm$ 3&13 $\pm$ 7&17 $\pm$ 8.5& 8 $\pm$ 4& 20  $\pm$ 14\\
SIMBA-a & ... & ...&... &... &5 $\pm$ 1&11  $\pm$ 2& 5 $\pm$ 1&11 $\pm$  2& ...\\
\hline
\end{tabular}
}
\end{table*}

\subsubsection{CCH lines}\label{sec:lte_cch}
We clearly detect all the components from the CCH ($N$=1-0) transition and around eight to nine (out of 11) hyperfine components of the CCH ($N$=3-2) transition depending on the source. In order to derive  excitation temperatures and column densities for the hyperfine structure of CCH, we used the Hyperfine fitting Structure (HfS) Tool in the CLASS software package. We derived the parameters separately for the data at 1mm and at 3mm because the HfS routine treats only one hyperfine transition at a time. By doing the analysis for the two transitions, we verified that we obtained the same physical parameters for the two transitions. The opacities derived show that most of the sources are moderately optically thick with values ranging from 0.8 to 2.2 at 1mm and from 0.6 to 3.3 at 3mm. The derived line widths, FWHM, and rest velocities, $V_{\text{lsr}}$, are similar to those derived from the Gaussian fits for most sources. The line widths may be larger for sources in which the line profile shows two components, such as the source SIMBA-a (see Appendix A). 

\indent The routine reads an input file containing the number of the components of the multiplet and their relative velocities, with respect to a chosen hyperfine component (here we chose the line at 262.004 GHz),  as well as their relative intensities. We used the predicted frequencies and intensities of the CCH lines for the N = 3-2 transition, taken from \citet{ziurys1982}.  In order to check the results and the predicted values, we used all the components of the multiplet, detected or not. The outputs given by the routine are the following:  p$_{1}$=T$_{mb}\times \tau$, the rest velocity V$_{\text{lsr}}$, the width of the lines $\Delta V$ (labelled here as FWHM) and the total opacity of the multiplet p4=$\tau$. From those parameters, we can extract the excitation temperature, $T_{\text{ex}}$, and the total column density of CCH, $N_{\text{tot}}$, for each source thanks to equations(~\ref{form:T}), (~\ref{form:Tex}), (\ref{form:J}), and (\ref{form:N}).

\begin{eqnarray}
\label{form:T}
J_{T_\text{ex}}=J_{T_\text{bg}}+\frac{\text{p}1}{\text{p}4}
\end{eqnarray}

where J(T), the intensity in units of temperature, is defined as:
\begin{eqnarray}
\label{form:J}
J(T)=\frac{h\nu}{k_{\text{B}}T}\frac{1}{e^{h\nu/k_{\text{B}}T}-1}
,\end{eqnarray}

where $h$, $\nu$, $k_{\text{B}}$ and T are the Planck constant, the frequency, the Boltzmann constant and the Temperature respectively.\\
This leads to :

\begin{eqnarray}
\label{form:Tex}
T_\text{ex}= \frac{h\nu /k_B}{\text{ln}(1+\frac{h\nu /k_B}{J_{T_\text{bg}}+\frac{\text{p}1}{\text{p}4}})}
.\end{eqnarray}

We then can calculate the total column density N$_{\text{CCH}}$:
\begin{eqnarray}
\label{form:N}
N_{\text{CCH}}=\frac{8\pi \nu ^3 \tau \Delta V Q(T_{\text{ex}})}{A_{ij}g_jc^3}\frac{e^{E_u/T_{\text{ex}}}}{e^{h c/\nu T_{\text{ex}}}-1}
,\end{eqnarray}

where  $E_u$, $A_{ij}$, $g_j$,   $Q(T_{\text{ex}})$, c and $T_{\text{ex}} $ are the upper level energy, the Einstein coefficient, the statistical weight of the upper level energy, partition function of the excitation temperature, the celerity and the excitation temperature, respectively. 

\indent The results of the hyperfine fit and of the LTE analysis are shown in Tables ~\ref{tab:lte-cch} and ~\ref{tab:results_LTE} respectively. The  excitation temperatures derived at 1mm and at 3mm are similar, as well as the total column densities. We thus averaged the results for the two transitions. The mean excitation temperature ranges from 5 to 17 K and the column density from  $6\times 10^{14}$cm$^{-2}$ to  $11\times 10^{14}$cm$^{-2}$.

\begin{table*}
\centering
\makegapedcells
\setcellgapes{2pt}
\caption{Results of the hyperfine structure fit of CCH for each source, at 1mm and 3mm.}
\label{tab:lte-cch}
\begin{tabular}{|c|c|cccc|}
\hline
\multirowthead{2}{Telescope}&\multirowthead{2}{Source} & $T_{\text{mb}}*\tau$ & V$_{\text{lsr}}$  & FWHM & \multirowthead{2}{$\tau_{\text{main}}$}  \\
&&[K]  & [km/s]  & [km/s] &    \\
\hline
\multirow{ 6}{*}{\rotatebox[origin=c]{90}{IRAM}}&CSO33& 13 $\pm$ 6  & 10.4 $\pm$  0.1 & 1.5 $\pm$  0.1 &2.2 \\
&FIR1a& 9 $\pm$ 5 & 11.5 $\pm$  0.1 & 1.2 $\pm$  0.1 & 0.8\\
&MMS9 & 8 $\pm$  4 & 11.4 $\pm$  0.1 & 1.0 $\pm$ 0.1& 1.0\\
&MMS5& 9 $\pm$ 7  & 11.0 $\pm$ 0.1 & 1.1 $\pm$  0.1 & 1.0\\
&MMS2 & 10 $\pm$ 4 & 11.1 $\pm$ 0.1 & 1.4 $\pm$ 0.1& 1.7\\
&CSO3 & 7 $\pm$ 5 & 11.0 $\pm$  0.1 & 1.1 $\pm$  0.1& 0.4\\
\hline
\multirow{7}{*}{\rotatebox[origin=c]{90}{Nobeyama}}&CSO33&6 $\pm$ 2 & 10.6 $\pm$ 0.2 & 1.3 $\pm$ 0.2 & 0.8 \\
&FIR1a& 8 $\pm$ 3&11.8 $\pm$  0.2 & 1.6 $\pm$ 0.2 & 0.6\\
&MMS9 & 9.0 $\pm$ 3.5 & 11.6 $\pm$ 0.2 & 1.1 $\pm$ 0.2 & 0.7  \\
&MMS5&12 $\pm$ 5& 11.2 $\pm$ 0.2 & 1.2 $\pm$ 0.2& 1.2 \\
&MMS2 &15 $\pm$ 6& 10.8 $\pm$ 0.2 & 1.4 $\pm$ 0.2 & 2.1  \\
&CSO3 & 13 $\pm$ 5& 10.9 $\pm$ 0.2 & 1.6 $\pm$ 0.2 & 1.5  \\
&SIMBA-a& 7 $\pm$ 3& 10.8 $\pm$ 0.2 & 1.7 $\pm$ 0.2 & 3.3 \\
\hline
\end{tabular}
\end{table*}

\subsection{OTF Map}

\begin{figure*}
\includegraphics[width=\linewidth]{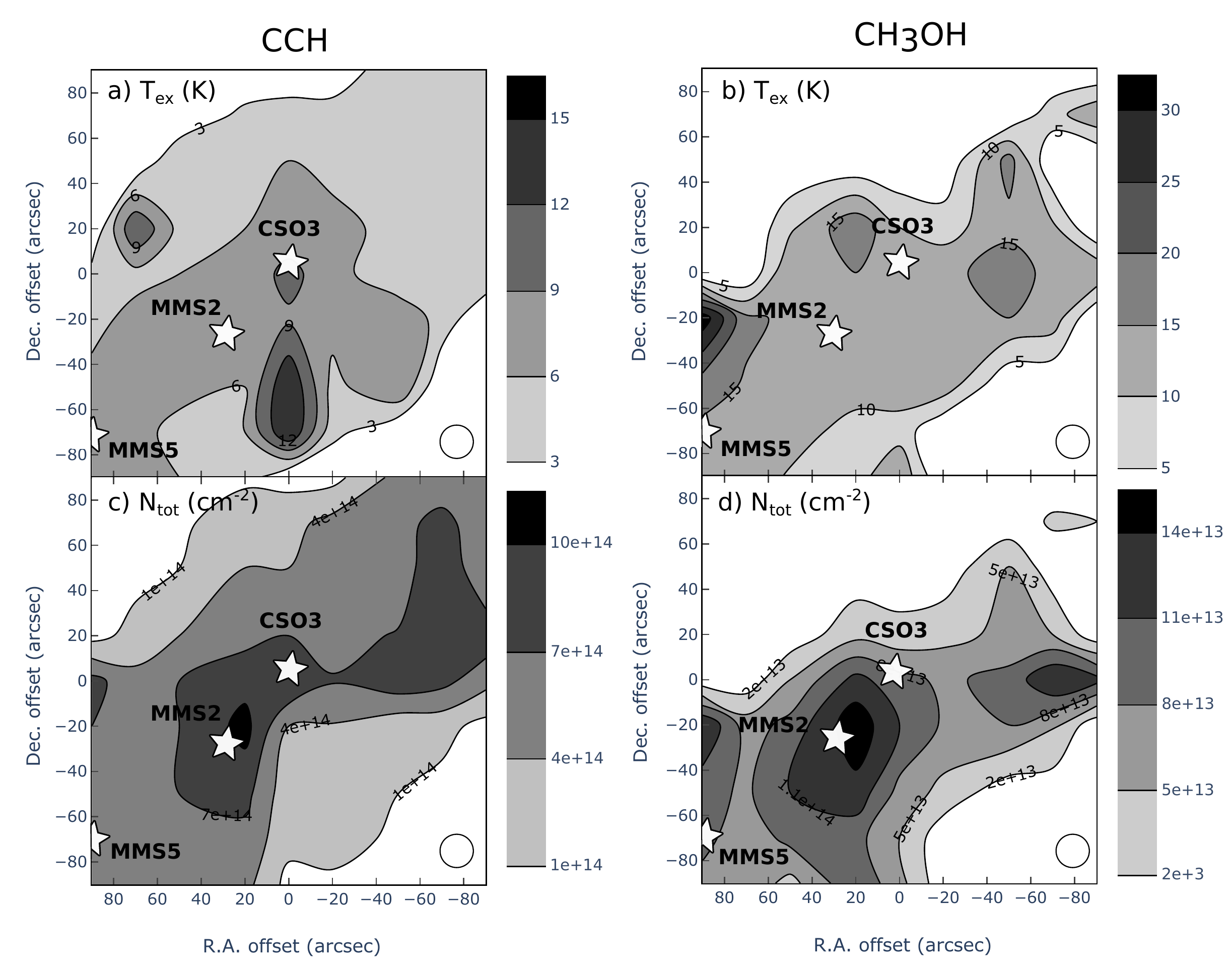}
\caption{LTE results for the OTF map. The size of the beam is shown by a white filled circle at the bottom right of each map. a) and b) Derived excitation temperature, Tex, of CCH and CH$_3$OH respectively. c) and d) Derived total column density,Ntot, of CCH and CH$_3$OH, respectively. }
\label{fig:maps_lte}
\end{figure*}

To derive the excitation temperatures and the column densities of each position of the map, we used the same methods as for the single-pointing. We took into account only positions with at least 3 detected lines of each molecules. The derived parameters are shown in Tab.~\ref{tab:lte_map} and the results in Fig.~\ref{fig:maps_lte}. The rotational temperatures range from 5 to 33 K and the column densities range from $2\times 10^{13}$ to $16\times 10^{13}$ cm$^{-2}$ for CH$_3$OH. For CCH, temperatures ranges from 4 to 15 K and column densities ranges from $1\times 10^{14}$ to $11\times 10^{14}$ cm$^{-2}$.

\FloatBarrier

\begin{table*}
    \centering
    \caption{Derived excitation temperatures and column densities of CCH and CH$_3$OH from the LTE analysis. Only positions with results (at least three lines) are shown here.}
    \label{tab:lte_map}
    \begin{tabular}{|cc|cc|cc|c|}
    \hline
    \multicolumn{2}{|c|}{Coordinates} & \multicolumn{2}{c|}{CH$_3$OH }& \multicolumn{2}{c|}{CCH} &Ratio \\
    \hline
      Offset R.A.   & Offset Dec. & T$_{\text{ex}}$ &N$_{\text{tot}}$ &T$_{\text{ex}}$& N$_{\text{tot}}$& \multirowthead{2}{[CCH]/[CH$_3$OH]}  \\
       $[$arcsec$]$ & [arcsec] &[K] &[$\times 10^{13}$ cm$^{-2}$] & [K]&[$\times 10^{14}$ cm$^{-2}$] & \\
      \hline
      +90 & -90 & 14 $\pm$ 10 &7 $\pm$ 3& 8.5 $\pm$ 0.5&5 $\pm$ 2 &7 $\pm$ 4\\
      +90 & -70 & 14 $\pm$ 2&  9 $\pm$ 2&7 $\pm$ 1&7  $\pm$ 2&8 $\pm$ 3\\
      +90 & -50 & 17 $\pm$ 3& 9 $\pm$ 2&7 $\pm$ 1&6  $\pm$ 2&7 $\pm$ 3\\
      +90 & -20 & 33 $\pm$ 11& 14 $\pm$ 3&5 $\pm$ 1& 7  $\pm$ 1 & 5$\pm$ 1\\
      +90 &  0 & ... & ... &4 $\pm$ 1 &8 $\pm$ 1& ... \\
      +70 & -90 & 15 $\pm$ 2&7 $\pm$ 1& 7 $\pm$ 1& 5 $\pm$ 2 &7 $\pm$ 3\\
      +70 & -70 & 14 $\pm$ 1&6 $\pm$ 1& 7 $\pm$ 1&5  $\pm$ 1 &8 $\pm$ 2\\
      +70 & -50 & 14.5 $\pm$ 1.5&7.5 $\pm$ 1.0&7 $\pm$ 1& 6 $\pm$  2&8 $\pm$ 3\\
      +70 & -20 & 16 $\pm$ 2&6.5 $\pm$ 1.0&7 $\pm$ 1& 6 $\pm$ 2 &9 $\pm$ 3\\
      +70 &  0 & ...& ...&5 $\pm$ 1& 5 $\pm$ 1 &... \\
      +70 & +20 & ...&... &12 $\pm$ 4&  2 $\pm$ 1& ...\\
      +50 & -90 & 9 $\pm$ 2&4 $\pm$ 1&6 $\pm$ 1& 5 $\pm$ 1 &12.5 $\pm$ 4\\
      +50 & -70 & 14 $\pm$ 1 &9 $\pm$ 1&5 $\pm$ 1&6  $\pm$ 1 &7 $\pm$ 1\\
      +50 & -50 & 13 $\pm$ 1&11 $\pm$ 2&7 $\pm$ 1& 7 $\pm$ 2 &6 $\pm$ 2\\
      +50 & -20 & 14 $\pm$ 1&9 $\pm$ 1&8 $\pm$ 1& 7 $\pm$ 3 &8 $\pm$ 3\\
      +50 &  0 & 15 $\pm$ 3&5 $\pm$ 1&6 $\pm$ 1&7  $\pm$ 1 &14 $\pm$ 2\\
      +50 & +50 & ...& ...&6 $\pm$ 1& 1.0 $\pm$  0.2& ...\\
      +20 & -90 &8 $\pm$ 2 &10.5 $\pm$ 1.0 &4.5 $\pm$ 1.0& 5 $\pm$ 1 &5 $\pm$ 1\\
      +20 & -70 &6.5 $\pm$ 1.0&9 $\pm$ 2&6 $\pm$ 1&6  $\pm$ 1 &7 $\pm$ 2\\
      +20 & -50 & 14 $\pm$ 1&13 $\pm$ 2 &6 $\pm$ 1&8  $\pm$ 2 &6 $\pm$ 2\\
      +20 & -20 & 15 $\pm$ 2&16 $\pm$ 2&8.5 $\pm$ 1.0& 11 $\pm$ 4 &7 $\pm$ 3\\
      +20 &  0 & 15 $\pm$ 2&12 $\pm$ 2&7 $\pm$ 1&9 $\pm$ 3  &7.5 $\pm$ 3\\
      +20 & +20 & ...& ...&6 $\pm$ 1& 6 $\pm$ 1 &... \\
      +20 & +50 & ...& ...&4 $\pm$ 1& 4 $\pm$ 1 & ...\\
      +20 & +70 & ...& ...&4 $\pm$ 1&4  $\pm$ 1 & ...\\
      0 & -90 & 14 $\pm$ 2&6 $\pm$ 1& ...& ... & ...\\
      0 & -70 & 8 $\pm$ 1 &5 $\pm$ 1&15 $\pm$ 6&2  $\pm$ 1 &4 $\pm$ 2\\
      0 & -50 &12.5 $\pm$ 1.0&7 $\pm$ 1&15 $\pm$ 4& 2 $\pm$ 2 &3 $\pm$ 2\\
      0 & -20 & 14 $\pm$1&11 $\pm$ 2&8.5 $\pm$ 1.0& 4 $\pm$ 2 &4 $\pm$ 2\\
      0 &  0 & 13 $\pm$ 1&9 $\pm$ 1&10 $\pm$ 2& 10 $\pm$ 5 &11 $\pm$ 6\\
      0 & +20 & 10 $\pm$ 2&3 $\pm$ 1&7 $\pm$ 1& 7 $\pm$ 2 &23 $\pm$ 10\\
      0 & +50 & ...& ...&6 $\pm$ 1& 4 $\pm$ 1 &... \\
      0 & +70 &... & ...&5 $\pm$ 1& 3 $\pm$ 1 &... \\
      -20 & -70 &... &... &4 $\pm$ 1& 3 $\pm$ 1 &... \\
      -20 & -50 & 13 $\pm$ 1&4 $\pm$ 1&5 $\pm$ 1&2  $\pm$ 1 &5 $\pm$ 3\\
      -20 & -20 & 12 $\pm$ 1&6 $\pm$ 1&...& ... & ...\\
      -20 &  0 & ...& ...&5.5 $\pm$ 1.0& 7 $\pm$ 1&... \\
      -20 & +20 &8 $\pm$ 2 &4 $\pm$ 1&7 $\pm$ 1&6  $\pm$ 2 &15 $\pm$ 6\\
      -20 & +50 &...&... &5 $\pm$ 1& 5  $\pm$ 1& ...\\
      -20 & +70 & ...&... &5 $\pm$ 1& 5 $\pm$ 1 &... \\
      -50 & -50 & ...&... &9 $\pm$ 1& 1.0 $\pm$ 0.2 &... \\
      -50 & -20 &15 $\pm$ 2&8 $\pm$ 1&8 $\pm$ 1&2 $\pm$ 1&2.5 $\pm$ 1.0\\
      -50 &  0 & 18 $\pm$ 2&9 $\pm$ 1&4 $\pm$ 1& 9 $\pm$ 1 &10 $\pm$ 2\\
      -50 & +20 & ...&... &5 $\pm$ 1&6  $\pm$ 1 &... \\
      -50 & +50 &17 $\pm$ 3 &5 $\pm$ 1 &5 $\pm$ 1& 5 $\pm$ 1 &10 $\pm$ 3\\
      -50 & +70 & ...&... &4 $\pm$ 1& 6 $\pm$ 1 & ...\\
      -50 & +90 & ...&... &6 $\pm$ 1&5  $\pm$ 1 &... \\
      -70 & -20 & 13 $\pm$ 1&5.5 $\pm$ 1&5 $\pm$ 1&2  $\pm$ 2 &4 $\pm$ 4\\
      -70 &  0 & 13 $\pm$1&14 $\pm$ 2&5 $\pm$ 1& 8 $\pm$ 1 &6 $\pm$ 1\\
      -70 & +20 &11 $\pm$ 1& 5 $\pm$ 1&4 $\pm$ 1& 9 $\pm$ 1 &18 $\pm$ 4\\
      -70 & +50 & ...& ...&4 $\pm$ 1& 10 $\pm$ 9 & ...\\
      -70 & +70 & 8 $\pm$ 2&2.5 $\pm$ 1&4 $\pm$ 1& 8 $\pm$ 1 &32 $\pm$ 13\\
      -70 & +90 &... & ...&5 $\pm$ 1& 5 $\pm$ 4 & ...\\
      -90 &  0 & 14.5 $\pm$ 1.0&12 $\pm$ 2&5.5 $\pm$ 1.0& 5  $\pm$ 1&4 $\pm$ 1\\
      -90 & +20 & ...& ...&4 $\pm$ 1& 9 $\pm$ 1 &...\\
      -90 & +50 & ...& ...&6 $\pm$ 1& 4 $\pm$ 1 & ...\\
      -90 & +70 & 14 $\pm$ 11&2 $\pm$ 1&5 $\pm$ 1& 5  $\pm$ 1 &25 $\pm$ 13\\
      -90 & +90 & ...& ...&6 $\pm$ 1& 4 $\pm$ 4 & ...\\
      \hline
    \end{tabular}
\end{table*}

\end{document}